\documentclass[12pt,notepaper,usenames,dvipsnames]{article}
\usepackage{jheppub} 
\usepackage{bm} 
\usepackage{bbm} 
\usepackage{mathrsfs} 
\usepackage{amscd} 
\allowdisplaybreaks[2] 

\usepackage{tikz, pgf} 
\usetikzlibrary{arrows, calc, shapes, shapes.arrows, decorations.markings, decorations.pathmorphing, cd}
    \tikzset{hasse/.style={circle, fill, inner sep=2pt}}
    \tikzset{bd/.style={circle, draw=black, inner sep=0pt, fill=black, minimum size=2mm}}
    \tikzset{wd/.style={circle, draw={red}, inner sep=0pt, fill=white, minimum size=2mm}}
    \tikzset{rd/.style={circle, draw=red, inner sep=0pt, fill=red, minimum size=2mm}}

\usepackage{xcolor}

\def\green#1{{\color{black!45!green}{#1}}}

\def\nn{\nonumber}
\def\ph{\phantom}
\newcommand{\bpm}{\begin{pmatrix}}
\newcommand{\epm}{\end{pmatrix}}
\newcommand{\bsm}{\begin{smallmatrix}}
\newcommand{\esm}{\end{smallmatrix}}
\newcommand{\bspm}{\left(\begin{smallmatrix}}
\newcommand{\espm}{\end{smallmatrix}\right)}

\def\del{{\partial}}
\def\til{\widetilde}

\def\^{\wedge}

\def\AutS{{\rm AutS}}
\def\AT{{\rm AT}}
\def\GalQ{{\rm GalQ}}
\def\CB{{\rm CB}}
\def\CBs{{\CB^*}}
\def\Re{{\rm Re}}

\def\GL{{\rm GL}}
\def\SL{{\rm SL}}
\def\PSL{{\rm PSL}}

\def\SU{{\rm SU}}
\def\PSU{{\rm PSU}}
\def\Sp{{\rm Sp}}

\def\SO{{\rm SO}}
\def\Spin{{\rm Spin}}
\def\u{\mathfrak{u}}
\def\su{\mathfrak{su}}
\def\sp{\mathfrak{sp}}
\def\so{\mathfrak{so}}
\def\fg{\mathfrak{g}}

\def\C{\mathbb{C}}

\def\P{\mathbb{P}}

\def\R{\mathbb{R}}
\def\Z{\mathbb{Z}}

\def\cF{\mathcal{F}}
\def\cH{\mathcal{H}}

\def\cN{\mathcal{N}}
\def\Nt{$\cN{=}2$}
\def\Nts{$\cN{=}2^*$}
\def\Nf{$\cN{=}4$}
\def\cS{\mathcal{S}}
\def\cV{\mathcal{V}}

\def\a{\alpha}
\def\ta{{\til\a}}
\def\b{\beta}
\def\g{\gamma}

\def\D{\Delta}

\def\z{\zeta}

\def\Th{\Theta}

\def\l{\lambda}

\def\tl{{\til\l}}
\def\L{\Lambda}
\def\tL{{\til\L}}
\def\r{\rho}
\def\s{\sigma}
\def\S{\Sigma}
\def\t{\tau}
\def\f{\phi}
\def\O{\Omega}
\def\w{\omega}

\title{Genus 2 Seiberg-Witten curves\\ 
for rank 2 $\mathcal{N}$=4 superYang-Mills theories}
\author[a]{Philip C. Argyres}
\author[b]{, Mario Martone}
\author[b,c]{, Zekai Yu}
\affiliation[a]{Physics Dept., U. Cincinnati, PO Box 210011, Cincinnati OH 45221, USA}
\affiliation[b]{Dept.\ of Mathematics, King’s College London, The Strand, London WC2R 2LS, UK}
\affiliation[c]{Qiuzhen College, Tsinghua University, Beijing, 10084, China}
\emailAdd{philip.argyres@gmail.com}
\emailAdd{mario.martone@kcl.ac.uk}
\emailAdd{yuzk23@mails.tsinghua.edu.cn}
\abstract{We determine new genus 2 Seiberg-Witten curves for four dimensional rank 2 absolute $\mathcal{N}$=4 superYang-Mills theories using the automorphism twist approach. 
The conformal manifolds of these curves agree with those predicted by S-duality orbits of global structures, and we use this to identify which of the two S-duality orbits of the $\so(5) \simeq \sp(4)$ superYang-Mills theory the genus-2 curve corresponds to.
We also compare the curves to earlier constructions of Seiberg-Witten curves for these theories as spectral curves of integrable systems.
These spectral curves have genus greater than the rank, and so only give a Coulomb branch geometry upon projection to a sublattice of the homology lattice of the curves.
We show how to determine the correct sublattice projection, and find that the integrable system curves do not apply to our theories.
}

\usepackage{todonotes}

\begin{document} 
\maketitle

\section{Introduction}

Although the low energy theories on the Coulomb branch (CB) moduli spaces of \Nf\ superYang-Mills (sYM) theories in 4 dimensions are by now very well-studied \cite{Seiberg:1994aj, Seiberg:1997ax, Donagi:1995cf, Witten:1997sc, DHoker:1997hut, DHoker:1998xad, DHoker:1998rfc, DHoker:1998zuv}, developments over the last decade or so refining the understanding of the global structures of these theories and their S-duality orbits \cite{Gaiotto:2010be, Aharony:2013hda, Bourget:2018ond, Argyres:2018wxu} have prompted a refinement of the description of their CB effective theories \cite{Tachikawa:2013kta, Argyres:2022kon, Closset:2023pmc}.
While perhaps the simplest description of \Nf\ CBs is as special Kahler (SK) orbifold geometries \cite{Seiberg:1997ax, Caorsi:2018zsq, Argyres:2019ngz, Argyres:2022kon, ABGLW23}, we focus here instead on their description in terms of Seiberg-Witten curves and one-forms \cite{Seiberg:1994rs, Seiberg:1994aj}.
The reasons for this are two-fold: firstly, the SW curve formulation can accommodate the deformation of these theories to non-conformal \Nts\ theories where the orbifold structure of the CB is lost; and secondly, the construction of SW curve descriptions provides a potential method for classifying all possible rank-2 CB geometries \cite{Argyres:2005pp, Argyres:2005wx, Argyres:2022lah, Argyres:2022puv, Argyres:2022fwy}, and thus to get model-independent constraints on the set of all possible rank-2 \Nt\ SCFTs.
In particular, one of the approaches to constructing rank-2 SW curves, advocated in \cite{Argyres:2022fwy}, is to take advantage of the relatively rigid form of genus-2 Riemann surfaces (RSs) with enlarged automorphism groups to construct SW curves and 1-forms in terms of discrete ``automorphism twist'' data.
This is the approach to constructing genus-2 SW curves and 1-forms for \Nf\ sYM theories which we will explore here.

In this latter regard, this paper can be seen as an exploration and development of the automorphism twist technique.
The interesting thing we find is that automorphism twists can be encoded and constructed algebraically in terms of a polynomial whose Galois group is the automorphism twist group of the SW curve and 1-form.
This association promises to apply more broadly to all ``isotrivial'' CB geometries \cite{Cecotti:2021ouq}.

There are some restrictions on the possible rank-2 CB geometry stemming from the assumption of a description in terms of a fibration of genus-2 RSs over $\C^2$  \cite{Argyres:2022lah}. 
They are that
\begin{itemize}
    \item[(i)] only geometries with principal Dirac pairing can occur, 
    \item[(ii)] there are no complex singularities on the CB, and
    \item[(iii)] the rank-2 geometries are ``non-split''.
\end{itemize}
The meaning of these restrictions has been described in  detail elsewhere, see, e.g., \cite{Argyres:2019ngz, Argyres:2022kon, Argyres:2022lah}.
In the case of the \Nf\ sYM CB geometries we study here, the first two restrictions give no constraint, while the third rules out one possible geometry.
In particular, there are five absolute rank-2 \Nf\ sYM theories with simple gauge algebra:  two $A_2$ theories, two $BC_2$ theories, and one $G_2$ theory.%
\footnote{We use the Dynkin notation for simple Lie algebras, $A_2 = \su(3)$, $BC_2 = \so(5) = \sp(4)$, and $G_2$.}
Each of these theories has an exactly marginal complex coupling constant and thus a 1-dimensional conformal manifold.  
We treat a family of theories parameterized by a connected conformal manifold as a single theory.
(This is also called an ``S-duality orbit" of theories in the literature.)
By ``absolute'' theory \cite{Freed:2012bs} we mean theories for which a (possibly twisted) partition function is defined on arbitrary compact spacetime manifolds without boundary.
Absolute sYM theories are ones whose lattice of electric and magnetic probe lines has principal Dirac pairing, i.e., the lattice comes with a symplectic pairing with invariant factors all equal to 1.%
\footnote{The SW curve description of scale-invariant CB geometries cannot determine the overall normalization of the Dirac pairing; this can only be determined with the addition of extra data, such as the spectrum of charged states or the \Nts\ deformation of the SW curve.}
So the restriction to principal Dirac pairing means that we are constructing curves for absolute \Nf\ theories.

The assumption that there are no complex singularities on the CB (or, equivalently, that the CB chiral ring is freely generated) does not eliminate any \Nf\ theories at rank 2.
For although there exist absolute \Nf\ theories with CBs with complex singularities, they all come from gauging discrete \Nf-preserving global symmetries of theories with freely-generated CB chiral rings \cite{Bourget:2018ond, Argyres:2018wxu}.
At rank 2 there is only one such possible discrete gauging, namely the theory related to the ``usual'' $A_2$ \Nf\ sYM theory by gauging its $\Z_2$ charge conjugation symmetry (which acts as an outer automorphism on the gauge algebra).  
But it so happens that gauging this $\Z_2$ results in a CB geometry which still has a freely-generated chiral ring.
In fact, the moduli space geometry of this ``$A_2\oplus\Z_2$'' \Nf\ sYM theory is identical to that of the $G_2$ \Nf\ sYM theory \cite{Argyres:2018wxu}.

Finally the restriction to non-split CB geometries eliminates one \Nf\ theory at rank 2.
A ``split'' CB geometry is one where there exists an EM duality frame in which the low energy effective $\u(1)^r$ gauge coupling, $\t^{ij}: {\rm CB} \to \cH_r$, is block diagonal.%
\footnote{Here $\cH_r$ is the rank-$r$ Seigel half-space of symmetric $r\times r$ complex matrices with positive definite imaginary part.}
Thus at rank 2 the only split CB geometries are those in which $\t^{ij}$ is diagonal.
In terms of the SW curve description, $\t^{ij}$ is the complex modulus of the RS.  
There are no smooth genus-2 RSs with diagonal complex modulus, so split CB geometries cannot be described by genus-2 SW curves.  
Instead, they are described by a singular limit of genus-2 RSs where the RS degenerates to a bouquet of two genus-1 RSs.
Among rank 2 \Nf\ sYM theories one of the two absolute $BC_2$ sYM theories is split \cite{Argyres:2019ngz, Argyres:2022fwy}.
There are two $BC_2$ theories since, by the analysis of \cite{Aharony:2013hda}, there are two distinct S-duality orbits of the global structures of the absolute theories.
The same analysis shows that there is only a single S-duality orbit of the $A_2$ \Nf\ global structures, and only a single $G_2$ global structure.

To summarize, from the analysis of rank 2 \Nf\ sYM theories we expect to find only 3 distinct CB geometries described by genus 2 SW curves, namely, those of the $A_2$ theory, the $G_2$ and $A_2\oplus\Z_2$ theories whose CB geometries are the same, and the non-split $BC_2$ theory.
In what follows we will construct precisely these three scale-invariant SK geometries in terms of genus-2 SW curves and 1-forms using the automorphism twist technique.
(The genus 2 $A_2$ curve was already constructed in \cite{Argyres:2022kon}.)
Although it is expected that these are the only simple rank 2 \Nf\ CB geometries (i.e., those which are isotrivial and have a single exactly marginal deformation) with genus 2 SW curve descriptions, we do not as yet have enough control over the classification of all scale-invariant genus 2 SW geometries to verify that this is indeed the case.

\paragraph{S-duality groups.}

The curves we derive all come in families depending on a single parameter, $c$, corresponding to the exactly marginal deformation parameter of the \Nf\ theory.
The $c$ parameter space is thus the conformal manifold of these theories, and so its complex geometry encodes the S-duality group of the theory.
We show how to compute this group in each case and show that it matches to the S-duality group expected from the field theory analysis \cite{Aharony:2013hda, Dorey:1996hx, Argyres:2006qr}.
In particular, as mentioned above, there are two absolute $BC_2$ theories --- which we will refer to as the $BC_2^{(1)}$ and $BC_2^{(2)}$ theories in what follows --- corresponding to the two S-duality orbits of their global structures, and, correspondingly, there are two distinct CB geometries, one split, and one non-split.
But it is not clear a priori which orbit corresponds to the non-split geometry.
By determining the S-duality group of the non-split geometry we are able to make that determination.

\paragraph{Comparison to other SW curves.}

Finally, other SW curves and one forms have been constructed to describe \Nf\ sYM theories.
The two main constructions are from integrable systems \cite{Donagi:1995cf} and from IIA/M theory brane configurations \cite{Witten:1997sc}.
In the integrable system approach these \Nf\ SW curves appear as the spectral curves of a Lax pair description of certain integrable system \cite{DHoker:1997hut, DHoker:1998xad, DHoker:1998rfc, DHoker:1998zuv, Bordner:1998xsa, Bordner:1998xs, Bordner:1998sw}, while in the IIA/M approach they are recast in ``class S'' as branched covers of a one-punctured torus or a 4-punctured sphere \cite{Gaiotto:2009we}.
In all these cases the resulting SW curves have genus greater than the rank of the CB, and a further projection to a sublattice of the homology lattice of the curves must be specified.
We briefly review this construction and then refine it by showing how to determine the correct sublattice projection. 
We then compute the Dirac pairing it induces on the charge lattice and the matrix of low energy couplings, and compare to the curves we construct.
As was pointed out in \cite{Argyres:2022kon}, for $A_2$ the known curves all describe only relative field theories, i.e., with non-principal Dirac pairings. 
We find that the known $G_2$ curve also describes a theory with non-principal Dirac pairing.
Of the $BC_2$ theories we find that the known curve describes only the split theory.
Thus the curves that we derive here do not correspond to any curves which have appeared previously in the literature.

\paragraph{Organization.}

In section \ref{sec2} we review SW curves and one forms, and the two repara\-metrization frames that we use to describe them. 
The orbifold structure of \Nf\ CBs is reviewed in section \ref{sec3}, from which we derive how the automorphism twist of the one-form basis is encoded by a polynomial whose Galois group is the orbifold group (the Weyl group of the gauge algebra).
We perform the explicit computation of the \Nf\ curves in section \ref{sec4}, and in section \ref{sec5} we show that our results do not depend on certain choices made in the computation.
Section \ref{sec6} then derives the conformal manifolds and S-duality groups of the curves, and section \ref{sec7} compares our curves to other curves proposed in the literature. 


\section{SW curves in canonical and automorphism frames}\label{sec2}

We briefly review the structure and description of Coulomb branches (CBs) of rank 2 \Nt\ SCFTs to set notation. 
For details, see \cite{Argyres:2022lah, Argyres:2022fwy}.

We assume that the CB has no complex singularity and hence it is $\C^2$ as a complex variety. 
Let $(u,v)\in\C^2$ be coordinates diagonalizing the complex scaling action of the SCFT on the CB,
with rational scaling dimensions
\begin{align}\label{DuvtopqR}
(\D_u, \D_v) \doteq (pR, qR) ,
\end{align}
where $p$, $q$ are positive coprime integers and $R$ is their rational common factor.  
We take $\D_u \leq \D_v$ without loss of generality.
A convenient scale-invariant coordinate on the $v\neq 0$ patch of CB is $w := u^{1/p}v^{-1/q}$.

CBs of absolute \Nf\ sYM theories correspond to an S-duality orbit of a choice of maximal line lattice \cite{Gaiotto:2010be, Aharony:2013hda, Argyres:2022kon}, so they have principal Dirac pairing. 
At rank 2, CB special K\"ahler geometry is thus described either by a genus 2 SW curve and 1-form, or by ``split'' genus 2 varieties --- a certain family of degenerate genus 2 RSs \cite{Argyres:2022lah}.
The methods explored here only apply to genus 2 SW curves and not to split genus 2 varieties.
The family of genus 2 Riemann surfaces can be described by a plane curve 
\begin{align}\label{swcurve}
    S: \qquad y^2  = \sum_{n=0}^6 c_n x^n , 
\end{align}
whose coefficients $c_n$ are locally holomorphic on the CB. 

The SW 1-form is given by
\begin{align} \label{sw1form}
   \L= \D_u u\, \O^u+\D_v v\, \O^v ,
\end{align} 
where $\{ \O^u, \O^v \}$ is a basis of holomorphic 1-forms on the curve satisfying the integrability condition
\begin{align}\label{inteq}
\del^v \O^u-\del^u \O^v = dg .
\end{align}
$g$ is a meromorphic function on the fiber, 
and $d$ is the exterior derivative along the fiber.

Define a canonical basis of one-forms by
\begin{align} 
   \w^0 :=\frac{xdx}{y} \qquad \w^1:=\frac{dx}{y},
\end{align}
and write 
\begin{align}
    \O^j = \sum_{k=0}^1 a^j_k \w^k.
\end{align}
with $j = 0, 1$ corresponding to $u_0 \doteq u$, $u_1 \doteq v$. 
The integrability condition \eqref{inteq} becomes a system of 8 coupled nonlinear holomorphic differential equations for the curve and 1-form coefficient functions, and for $g$;  they are recorded in \cite{Argyres:2022fwy}.

The CB geometry is singular along a codimension-1 subvariety $\cV\subset\CB$, where the low energy theories include extra massless modes.
$\cV$ is, by scale invariance of the UV theory, invariant under the complex scaling symmetry action, so can only be a union of $\{u=0\}$, $\{v=0\}$, or $\{u^q \propto v^p\}$ components.
The linear system of homology 1-cycle lattices $H_1(S,\Z) \simeq \Z^4$ over the CB has a principal symplectic pairing given by the intersection product on $H_1$.
The electric-magnetic (EM) monodromy group is the image of the associated monodromy map, $\pi_1(\CBs) \to \Sp(4,\Z)$, of the linear system, where $\CBs = \CB \setminus \cV$ denotes the smooth part of the CB.


Finally, this description of SW curves and holomorphic 1-forms is redundant.
There is a coordinate reparameterization that affects both the coefficients of the curve as well as the 1-forms,
\begin{align}\label{repar}
x &\to \frac{G^0_0 x+G^0_1}{G^1_0 x+G^1_1}, & 
y &\to \frac{y}{(G^1_0 x+G^1_1)^3}, &
\w^i & \to \w^j \, (\det G) \, G_j^i,
\end{align}
for $G^i_j \in \GL(2,\C)/\Z_3$. 
It will be convenient to define 
\begin{align}\label{AdetGG}
A^i_j \doteq (\det G) \, G^i_j \in \GL(2,\C)/\Z_3 .
\end{align}

\paragraph{Canonical frame.}

The reparameterization freedom is completely fixed by transforming to a frame in which the holomorphic 1-forms are 
\begin{align} 
\O^u=\w^0 \qquad \O^v=\w^1.
\end{align}
This choice is called the canonical frame.
In canonical frame the $c_n$ curve coefficients are meromorphic (i.e., single-valued) functions on the CB.
Furthermore, in this frame the CB singularities $\cV\subset\CB$ correspond to the locus where the curve degenerates.
This is the locus where the discriminant with respect to $x$ of the right side of \eqref{swcurve} vanishes.
The EM monodromies are computed by dragging a basis of homology 1-cycles around closed paths in the CB which link $\cV$.

\paragraph{Automorphism frame.}

An alternative way of fixing the reparameterization invariance is to use the $\GL(2,\C)/\Z_3$ coordinate transformations to put the coefficients of the curve in a standard form, as described in \cite{Argyres:2022fwy}.
In such a frame where the curve $S$ does not vary over the CB --- and so has a fixed $\t$ --- the singularities, $\cV$, on the CB are not detected by degenerations of $S$ (there are none).
Instead, one imposes the EM monodromy ``by hand'' by identifying $S$ with itself up to an \emph{automorphism twist},
\begin{align}\label{aut twist}
   \a: \pi_1(\CBs)\to \AutS.
\end{align}
The automorphism group of $S$ is the finite subgroup $\AutS \subset \GL(2,\C)/\Z_3$ of reparameterizations which do not change the coefficients of the curve describing $S$.
The automorphism groups of genus 2 RSs are reviewed in \cite{Argyres:2022fwy}.
We call the image of $\a$ the ``automorphism twist subgroup'' 
\begin{align}\label{AT def}
    {\rm im}(\a) \doteq \AT \subset \AutS .
\end{align}

The automorphisms act on 1-cycles, inducing a representation of $\AutS$ in $\Sp(4,\Z)$.
So in the automorphism frame the EM monodromy group is identified with the image of the automorphism twist subgroup of $\AutS$ in $\Sp(4,\Z)$ under this representation.

Also, by \eqref{repar}, an automorphism $A\in\AutS \subset \GL(2,\C)/\Z_3$ acts on the 1-form coefficient matrix $a^i_j$ as
\begin{align}\label{amap}
A\in \AutS : a^i_j \mapsto A^k_j a^i_k ,
\end{align}
where $A^k_j$ is given by \eqref{AdetGG}.
Thus the associated automorphism twist monodromies are reflected in ``jumps'' (i.e., multi-valuedness) of the 1-form coefficient matrix by \eqref{amap} upon traversing closed paths in $\pi_1(\CBs)$.
This implies that the automorphism twist subgroup is the group of monodromies of the sheets of $a^i_j$ as a matrix of functions branched over the CB.

\section{Automorphism twists for \Nf\ CBs}\label{sec3}

\subsection{\Nf\ orbifold CBs}

We are interested in constructing genus 2 SW curves and 1-forms for absolute \Nf\ sYM theories with rank 2 simple gauge algebras $B_2 = \so(5) = C_2=\sp(4)$, and $G_2$.
(The $A_2 = \su(3)$ case was constructed in \cite{Argyres:2022fwy}.)
Denote the Weyl group of a simple Lie algebra $\fg$ by $W(\fg)$.
Since $B_2=C_2$ we denote it simply by $BC_2$, and the corresponding Weyl group by $W(BC_2)$.
Weyl groups are finite irreducible real crystallographic reflection groups, so $W(BC_2)$ and $W(G_2)$ are generated by reflections and act irreducibly on $\R^2$. 

The full moduli space of a $\fg=BC_2$ or $G_2$ \Nf\ sYM theory is the orbifold $\C^6/(W(\fg)\otimes_\R \C^3)$.
The whole \Nf\ Coulomb branch geometry can be reconstructed from an \Nt\ slice \cite{Argyres:2019yyb}.
So we focus on an \Nt\ CB slice given by $\CB = \C^2/W_\C(\fg)$, where $W_\C(\fg)=W(\fg)\otimes_\R \C$ is the complexification of the Weyl group. 
Denote by $(\tL_0,\tL_1) \in \C^2$ some basis of complex linear coordinates on which the Weyl group acts faithfully by linear transformations $\r: W_\C(\fg) \to \GL(2,\C)$.

\Nf\ supersymmetry implies that the special K\"ahler metric on CB is flat, the coupling matrix $\t^{ij}$ is constant, and suitable linear combinations of the $\tL_j$ are the special coordinates \cite{Argyres:2019yyb}.
The global complex $(u,v)$ coordinates on CB are a homogeneous basis of $W_\C(\fg)$-invariant polynomials in the $\tL_j$.
Since the special (and dual-special) coordinates are given by periods of the SW 1-form, by \eqref{sw1form} they are expressed in terms of the 1-form matrix coefficients and the periods of the canonical 1-form basis by
\begin{align}\label{Lperiod}
\oint \L = ( \D_u \, u\, a^u_j + \D_v\, v\, a^v_j ) \oint \w^j .
\end{align}

The pre-image in $\C^2$ of the singular locus $\cV\subset\CB$ are the points fixed by the action on $\C^2$ of at least one non-identity element of $W_\C(\fg)$.
Since Weyl groups are real reflection groups, the codimension-1 components of $\cV$ are fixed by some reflections $r \in W_\C(\fg)$.
These elements are thus associated to simple closed paths linking the codimension-1 components of $\cV$, and so generate a surjective representation, $\pi_1(\CBs) \to \r(W_\C(\fg)) \subset \GL(2,\C)$, of the fundamental group of the CB onto the Weyl group.
The special K\"ahler structure of the CB is an identification of the EM duality monodromy with this orbifold monodromy,
and thus gives an isomorphism of the EM monodromy group with the Weyl group. 

\subsection{\Nf\ automorphism frame curves}

Since $\t$ is constant on the CB of \Nf\ sYM theories, so is the SW Riemann surface.
Thus we can choose a constant curve $S$, i.e., one whose $c_n$ coefficients are constants on the CB.
In this case the EM duality monodromies around $\cV\subset\CB$ appear as multi-valuedness of the $a^i_j$ 1-form coefficients matrix elements over the CB, reflecting automorphism twists.

Since the image of the automorphism twist in the automorphism group of the curve, $\a(\AutS)$, is isomorphic to the EM duality group which is isomorphic to the orbifold group, it follows that the curve must have an automorphism group which contains the Weyl group as an abstract group,
\begin{align}\label{AutS>W}
   \AutS \supset W(\fg)  .
\end{align}
Furthermore, since \Nf\ sYM theories have a marginal coupling, the curves must actually be selected from a 1-parameter family of curves whose automorphism group satisfies \eqref{AutS>W}.

A glance at table 1 of \cite{Argyres:2022fwy}, which lists the possible automorphism groups of genus 2 RSs, shows that there is only one possible family of curves with the right automorphism groups to correspond to each rank 2 \Nf\ sYM theory:
\begin{align}\label{table}
\begin{array}{lll}
\fg & S(c) & \AutS \\
\hline
A_2 & y^2 = x^6+c x^3+1 & D_6 \supset W(A_2) \\
G_2 & y^2 = x^6+c x^3+1 & D_6 \simeq W(G_2) \\
BC_2\quad & y^2 = (x^2-1)(x^4+c x^2+1)\quad & D_4 \simeq W(BC_2) 
\end{array} \qquad .
\end{align}
Here $S(c)$ is a family of curves in automorphism frame and $\AutS$ is the automorphism group of the curve at generic value of the parameter $c$.
(The enhancements of the automorphism group at special values of $c$ will be discussed in section \ref{sec6}.)
These identifications follow from the group inclusions shown and the fact that there are no other 1-or-more-parameter families of curves with automorphism groups that include the Weyl groups. 
Here $D_n$ is the order-$2n$ dihedral group (a.k.a.\ $D_{2n}$).
$c$ is the free parameter of the curve $S(c)$, and since there is only one, it is identified with a complex coordinate on the conformal manifold of the corresponding theory, i.e., it is the marginal coupling of the \Nf\ sYM theory.

Our task is then to calculate the multi-valued 1-form coefficient matrix $a^i_j(u,v)$ branched over the $\CB$ with monodromies specified by \eqref{amap} for $A \in W(\fg) \subset \AutS$.
(Once that is done, we can invert \eqref{repar} to get back to the more familiar and single-valued canonical frame description.)

Since the automorphism frame curves are constant on the CB, so are the canonical basis holomorphic 1-forms, $\w^j$.
The special K\"ahler integrability condition \eqref{inteq} then reads $(\del^v a^u_j-\del^u a^v_j) \w^j = dg$. By the fact that $\w^j$ are linearly independent and not exact on $S(c)$, $dg$ must vanish, giving the pair of holomorphic linear differential equations
\begin{align}\label{inteq2}
   \del^v a^u_j-\del^u a^v_j = 0  .
\end{align}
Locally on the CB, these equations have infinitely many solutions.
The challenge is to construct solutions with the required monodromies.

The integrability condition and analysis of the monodromies can be simplified using the scaling symmetry of the CB.
Since the coefficients of $S(c)$ are constants, they are dimensionless under the scaling.
This implies the curve $(x,y)$ coordinates, and so also the canonical 1-forms, $\w^j$, are dimensionless in this frame.
Because its periods compute masses, the SW 1-form \eqref{sw1form} has scaling dimension 1, and so the 1-form coefficient matrix elements have dimensions $\D(a^u_j) = 1-\D_u$, and $\D(a^v_j) = 1-\D_v$. 
Define the dimensionless CB coordinate and quantities
\begin{align}\label{scaling}
    w &\doteq u^{1/p}v^{-1/q} , &
    \a^u_j(w) &\doteq v^{(\D_u-1)/\D_v} \, a^u_j , &
    \a^v_j(w) &\doteq v^{(\D_v-1)/\D_v} \, a^v_j .
\end{align}
The simplified integrability equations \eqref{inteq2} become the ordinary differential equations
\begin{align}\label{inteq3}
   0 = \D_u\, w\, (\a^u_j)' + \D_v\, w^{1-p}\, (\a^v_j)'
   + (\D_u-1)\, p\,\a^u_j .
\end{align}
Here the prime denotes derivative with respect to $w$.
The image in the $w$-plane of codimension-1 singular subvarieties of the CB are now a collection of points over which the $\a^i_j$ are branched with monodromies still given by \eqref{amap}.

Recall that special coordinates are given by the periods \eqref{Lperiod}.
Since in the automorphism frame the $\w^i$ are dimensionless and constant over the CB, their periods are pure numbers.
So the special coordinates are certain linear combinations of 
\begin{align}\label{Ltoa}
\L_j \doteq \D_u \, u\, a^u_j + \D_v \, v\, a^v_j , 
\qquad j \in \{0,1\} .
\end{align}
As noted earlier, these are a basis of flat complex coordinates on $\C^2$ which are acted on by the orbifold group $W(\fg)$.
Define dimensionless flat coordinates $\l_j(w) \doteq  v^{-1/\D_v}\,\L_j$, so that \eqref{Ltoa} becomes
\begin{align}\label{ltoa}
   \l_i = \D_u \, w^p \, \a^u_i + \D_v \, \a^v_i .
\end{align}
Then the special K\"ahler integrability condition \eqref{inteq3} can be rewritten as the pair of equations
\begin{align}\label{simpint}
   \l_j' = p \, w^{p-1} \, \a^u_j , \qquad j\in \{0,1\},
\end{align}
and \eqref{ltoa} and \eqref{simpint} can be inverted to give
\begin{align}\label{atol}
\a^u_j &= \frac1p w^{1-p} \l_j'  , &
\a^v_j &= \frac1{Rq} ( \l_j - R \, w \, \l_j' ) .
\end{align}
(Recall the definition \eqref{DuvtopqR} of $p$, $q$, and $R$.)
This reformulation will prove to be very useful for imposing the required monodromies on the solutions. 

\subsection{\Nf\ automorphism twists}

The orbifold description of the CB implies an algebraic relationship between the special coordinates and the global holomorphic $(u,v)$ coordinates on the CB.
In particular, the special coordinates are certain linear combinations of the $\tL_j\in\C^2$ flat coordinates on the orbifold, and the $(u,v)$ coordinates are irreducible Weyl-invariant homogeneous polynomials in the $\tL_j$.
This implies that $\tL_j$ are related to $(u,v)$ via a polynomial relation
\begin{align}\label{Q0}
Q(\tL, u,v) = 0,
\end{align}
where $Q$ is an irreducible minimal degree polynomial in $\tL$ whose Galois group is the orbifold group $W(\fg)$.
This is because \eqref{Q0} implies that $(u,v)$ are polynomial combinations%
\footnote{More precisely, \eqref{Q0} only implies that certain monomials in $u$, $v$ are polynomial combinations of  the $\tL_a$.  
In the case of reflection groups, the Shephard Todd theorem \cite{shephard_todd_1954} assures the existence of a polynomial for which $u$ and $v$ are separately homogeneous polynomials in the $\tL_a$.}
of the roots, $\tL_a$, of $Q(\tL)=0$ which are invariant under the symmetry group of those roots.

This symmetry group is the Galois group, $\GalQ$, of the field extension of the field of rational functions%
\footnote{I.e., the field of fractions of the polynomial ring $\C[u,v]$.} 
of $u$ and $v$ associated to the polynomial $Q$.
$\GalQ$ can be computed as the group of deck transformations on roots thought of as sheets of a degree $|Q|$ covering of the $(u,v)$ plane.
Here $|Q|$ denotes the degree of $Q$ in $\tL$.
Concretely, $\GalQ$ is the group generated by permutations of sheets $\tL_a(u,v)$ when going around given branching loci in the $(u,v)$ plane --- i.e., around co-dimension 1 components of $\cV$ in the CB.  
These deck transformations then furnish a faithful representation 
\begin{align}\label{deck rep}
D: \GalQ \to \cS_{|Q|}, 
\end{align}
where $\cS_n$ is the permutation group on $n$ objects.
Note that since $|Q|$ is generally larger than 2, the roots $\tL_a$, $a\in\{1,\ldots,|Q|\}$, form an over-complete linear basis of flat coordinates on the CB.  
A basis of special coordinates, $\L_j$, $j\in\{0,1\}$, on the CB is thus some linear combination of the roots of $Q$,
\begin{align}\label{LCtL}
\L_j = C^a_j \tL_a,
\end{align}
where $C^a_j$ is a constant $2\times |Q|$ matrix.

Since we ultimately want to solve for the 1-form coefficient matrix elements $a^i_j$, rewrite \eqref{LCtL} using the integrability relations \eqref{atol}.
In particular, by going to scale invariant coordinates
\begin{align}\label{tldef}
\l_j(w) & \doteq v^{-1/\D_v} \L_j, & 
\tl_a(w) & \doteq v^{-1/\D_v} \tL_a,
\end{align}
and defining the combinations
\begin{align}\label{attolt}
\ta^u_a &\doteq \frac1p w^{1-p} \tl_a'  , &
\ta^v_a &\doteq \frac1{Rq} ( \tl_a - R \, w \, \tl_a' ) ,
\end{align}
following \eqref{atol}, the relation \eqref{LCtL} between special coordinates and roots of $Q$ becomes the pair of relations
\begin{align}\label{aCta}
\a^i_j &= C^a_j \, \ta^i_a  .
\end{align}

The $a^k_j$ and therefore the $\a^k_j$ transform under automorphism twists via the $\GL(2,\C)/\Z_3$ action \eqref{amap}
\begin{align}\label{amonod}
 \a^i_j \overset{\g}{\leadsto} [A(\g)]^k_j \, \a^i_k,
\end{align}
where $\overset{\g}{\leadsto}$ denotes the monodromy upon analytically continuing along a path $\g\in\pi_1(\CBs)$.
The $\tL_a$ and therefore the $\tl_a$ transform under $\GalQ$ via the $|Q|$-dimensional representation \eqref{deck rep} of deck transformations.
Since the $\ta^i_a$ are defined as linear combinations of $\tl_a$ and $\tl_a'$, they transform in the same way,
\begin{align}\label{atmonod}
\ta^i_a \overset{\g}{\leadsto} \ta^i_b \, [D(\g)]^b_a .
\end{align}
Combining \eqref{amonod}, \eqref{atmonod}, and \eqref{aCta} gives
\begin{align}\label{intertwiner}
   [A(\g)]^k_j \, C^a_k &= C^b_j \, [D(\g)]^a_b 
   \qquad \text{for all $\g\in\pi_1(\CBs)$.} 
\end{align}

$A$ is a representation of the group of automorphism twists. It is the same as the group of EM duality monodromies and the orbifold Weyl group $W(\fg)$, which is isomorphic to $\GalQ$, of which $D$ gives a representation.  
Thus $A(\g)\in \GL(2,\Z)/\Z_3$ and $D(\g)\in \cS_{|Q|}$ are different representations of the same abstract group, and \eqref{intertwiner} shows that the matrix $C$ is an intertwiner between these representations, i.e., it is a $\GalQ$-equivariant homomorphism.
Given a monodromy representation of $\GalQ$ on $\CBs$, it is a matter of simple linear algebra to solve for the intertwiner $C$ between these representations.

The 1-form coefficient matrix $\a^i_j$ is determined by \eqref{aCta} in terms of $C$ and the $\ta^i_a$.
The $\ta^i_a$ are, in turn, defined in \eqref{attolt} in terms of $\tl_a$, which are the (dimensionless) roots of the $Q$ polynomial.
They are thus some algebraic functions of the dimensionless coordinate, $w$, on the CB for which there are generally no simple expressions.
But there is no need for a more explicit description, since, upon transforming from automorphism frame to canonical frame by taking $G=\det(a)^{1/3} a^{-1}$ in \eqref{repar}, the curve coefficients will automatically be $\GalQ$-invariant combinations of the $\tl_a$, and so will be rational functions of $(u,v)$.

In general, finding a polynomial with a specified Galois group can be a difficult algebraic exercise.
But in our case the task is greatly simplified by the fact that the relevant Galois groups are small --- $W(A_2)\simeq \cS_3$, $W(BC_2)\simeq D_4$, or $W(G_2) \simeq D_6$ --- together with the scale symmetry on the CB.
The scale symmetry implies that the linear coordinate $\tL$ has dimension 1, while $u$ and $v$ have fixed dimensions $\D_u$ and $\D_v$ given by the exponents (plus 1) of the Weyl groups.
It then turns out that the minimal degree homogeneous irreducible polynomial in $(\tL, u, v)$ automatically gives the desired $Q$, as we explain in the next section.

\section{Calculation of \Nf\ curves}\label{sec4}

In this section we illustrate the abstract discussion of the last section in three examples, namely, \Nf\ gauge theories with gauge algebras $A_2$, $G_2$, and non-split $BC_2$. 

\subsection{$A_2$ curve}

We start by reviewing the automorphism twist derivation of the $A_2$ \Nf\ sYM curve from \cite{Argyres:2022fwy}.
This will be useful for comparison to the $G_2$ and $BC_2$ curves, and will also be needed for the discussion of S-duality groups in section \ref{sec6}.

$W(A_2)=\cS_3$, so, from \eqref{table}, the automorphism frame curve is
\begin{align}\label{S A2}
   S(c) : \qquad y^2=x^6+cx^3+1
\end{align}
where $c$ is constant along the CB. This curve has automorphism group $\AutS = D_6 \supset \cS_3$ at generic values of $c$.

The $A_2$ CB scaling dimensions are $\D_u=2$ and $\D_v=3$ so one has a coprime pair $(p=2,q=3)$ with $R=1$.
From the field theory, $u,v$ are a basis of invariant polynomials of the $W(A_2)$ action on the (dimension 1) special coordinates $\tL \in \C^2$ which can be thought of as the complexification of a Cartan subalgebra of the gauge algebra $A_2$.
There is thus a polynomial $Q(\tL)$ with coefficients in $\C[u,v]$ whose Galois group is $\GalQ = W(A_2)$. 
The lowest-order scaling-homogeneous polynomial is simply
\begin{align}\label{Q A2}
    Q = \tL^3 -3 u\tL +2 v,
\end{align}
whose discriminant is $\sim u^3 - v^2$.
(We have used the freedom to rescale $u$ and $v$ to set the coefficients of $Q$ to convenient values.)
The discriminant vanishes at locus corresponding to the fixed point locus of the orbifold action, indicating that $\GalQ=W(A_2)$.
This can be checked directly by computing the Galois group of $Q$, e.g., \cite{Conrad:Gal34}.

The dimensionless version of $Q$ reads
\begin{align}\label{Th A2}
   \Th &= \tl^3 -3 w^2\tl + 2 ,&
   w &\doteq u^{1/2} v^{-1/3},
\end{align}
and $\tl \doteq \tL v^{-1/3}$. 
$\Th(\tl)=0$ is a 3-sheeted cover of the $w$-plane branched over $w^6=1$.
Label the three sheets $\tl_a$ for $a\in \Z_3$.
\eqref{Th A2} and \eqref{attolt} then give three solutions to the integrability equation,
\begin{align}\label{ta A2}
    \ta^u_a &\doteq \tl_a (\tl_a^2-w^2)^{-1}, &
    \ta^v_a &\doteq -\frac23(\tl_a^2-w^2)^{-1} .
\end{align}

The deck transformations of the $\tl_a$ are easily determined: upon traversing a small counterclockwise loop around the $w^6=1$ branch points, the sheets undergo the monodromies
\begin{align}\label{lt A2}
    \tl_a &\mapsto \tl_a, &
    \tl_{a\pm 1} &\mapsto \tl_{a \mp1}, &
    &\text{around}& w^2 &= \z^{-a}, 
\end{align}
where $\z$ is a primitive third root of unity.
The $\ta^k_a$ enjoy the same monodromies by virtue of \eqref{ta A2}.
They thus transform under the 3-dimensional deck transformation representation of $\GalQ \simeq W(A_2)$.

The 1-form coefficient matrix elements $\a^k_j$ are linear combinations of the $\ta^k_a$ which transform as a 2-dimensional representation given by the $\GL(2,\C)/\Z_3$ action of the automorphism twist group, which is $W(A_2) \subset \AutS$. 
Since $W(A_2)$ is generated by its reflections which fix the $u^3=v^2$ discriminant locus, i.e., the $w^6=1$ branch points, the $\a^k_j$ monodromies about those branch points must be generators of $\Z_2$ subgroups of $W(A_2) \subset \AutS$. 
There are three possible $\Z_2$ generators in $\AutS$, given as elements of the $\GL(2,\C)/\Z_3$ reparameterizations fixing the curve \eqref{S A2},%
\footnote{This representation of the automorphism groups of all genus 2 curves was computed in \cite{Argyres:2022fwy}.} 
which, via \eqref{amap} and \eqref{AdetGG}, give the possible monodromies
\begin{align}\label{AT A2}
    A_n: \bpm \a^k_0\\ \a^k_1 \epm &\mapsto 
    - \bpm 0 & \z^{-n} \\ \z^n & 0 \epm 
    \bpm \a^k_0 \\ \a^k_1 \epm
\end{align}
around $w^6=1$, where $n\in \Z_3$.

So we look for constant $2\times3$ matrices $C$ in \eqref{aCta} such that $\a^k_j$ have $A_n$ monodromies for some assignment of $n\in\Z_3$ to each $w^6=1$ branch point.
Elementary algebra shows that there are two possible assignments of $A_n$'s which give nondegenerate solutions for $C$ unique up to overall normalization, namely
\begin{align}\label{C A2}
    C = \bpm -1 & -\z^{2n} & -\z^n \\ \z^n & \z^{2n} & 1 \epm
    \ \ \text{for}\ n = \pm 1, 
\end{align}
if $w^2=\z^m$ is assigned monodromy $A_{n(m+1)}$.
We can take $n=1$ without loss of generality,  reflecting the cube root of unity factor ambiguity in our choice of $w^2 \doteq u v^{-2/3}$ as scale-invariant coordinate.
This gives the solution --- unique up to overall normalization and overall permutation of the $\l_n$ --- for the automorphism frame 1-form basis coefficient matrix
\begin{align}\label{form A2}
4(1-w^6) \a^u_j &= 
(-\z)^j 
(\tl_0 {-} \tl_1) (\tl_0 \tl_1 {+} w^2) (\tl_2^2 {-} w^2)
+ (-\z)^{1-j} 
(\tl_1 {-} \tl_2) (\tl_1 \tl_2 {+} w^2) (\tl_0^2 {-} w^2)
, \nn\\
6 (w^6-1)\a^v_j &= 
(-\z)^j 
(\tl_0^2 {-} \tl_1^2) (\tl_2^2 {-} w^2)
+ (-\z)^{1-j}
(\tl_1^2 {-} \tl_2^2) (\tl_0^2 {-} w^2)
. 
\end{align}
Reverting to the CB scaling coordinates using \eqref{scaling}, this matrix is
\begin{align}\label{amat A2}
a = \bpm  
v^{-1/3} \a^u_0 & v^{-1/3} \a^u_1 \\
 v^{-2/3} \a^v_0 & v^{-2/3} \a^v_1 
\epm .
\end{align}

Finally, it follows from \eqref{amap} and \eqref{AdetGG} that this automorphism frame solution is transformed to canonical frame curve by acting on the automorphism frame curve \eqref{S A2} with the $\GL(2,\C)/\Z_3$ reparameterization given by $G= a^{-1} (\det a)^{1/3}$.
A computer-aided calculation results in the curve
\begin{align}\label{curve A2}
y^2 =  \frac1{3^2 2^6(u^3-v^2)} \Bigl\{
& \ \ \ \ x^6 \cdot 2^6 
\ \ \, \left[(c{+}2) u^3 - 4 v^2\right]  
\nn\\
&+ x^5 \cdot 3^2 2^6 
\ (c{-}2) u^2 v
\nn\\
&+ x^4 \cdot 3^3 2^4 \left[(c{-}10) u^3 + 4c v^2\right] u 
\nn\\
&+ x^3 \cdot 3^3 2^5 \left[(3c{-}10) u^3 + 2c v^2\right] v 
\nn\\
&+ x^2 \cdot 3^5 2^2 \left[(c{+}10) u^3 + 4 (c{-}5) v^2\right] u^2 
\nn\\
&+ x \, \cdot\, 3^6 2 
\ \left[(c{+}6) u^3 - 8 v^2\right] u v 
\nn\\
&+ \ \ \ \ \ 3^6 
\ \ \, \left[(c{-}2) u^6 + 16 u^3 v^2 - 16 v^4\right]
\Bigr\} ,
\end{align}
derived first in \cite{Argyres:2022fwy}.
The curve has discriminant $\propto (c^2-4)^3 (u^3 - v^2)^5$, and the $c \to \pm 2$ curve degenerations are weak coupling limits.

\subsection{$G_2$ curve}

Since $W(G_2)=D_6$, by \eqref{table} the $G_2$ automorphism frame curve must be
\begin{align}\label{S G2}
   S(c): \qquad y^2=x^6+cx^3+1
\end{align}
where $c$ is constant along the CB. 
This is the same as the $A_2$ automorphism frame curve.

The $G_2$ CB scaling dimensions are $\D_u=2$ and $\D_v=6$, so one has coprime pair $(p,q)=(1,3)$ with $R=2$.
The lowest-order scaling-homogeneous polynomial $Q(\tL)$ with coefficients polynomial in $(u,v)$ is $Q=\tL^6+A u\tL^4+B u^2\tL^2 +(Cv+Du^3)$ where $A, B, C, D$ are undetermined complex coefficients.  
By rescaling $u$ and $v$ and by shifting $v$ by a multiple of $u^3$, the $A$ and $C$ coefficients can be fixed to any convenient values, and we can set $D=0$.  
The $B$ coefficient, however, must then be chosen so that $\GalQ = W(G_2) \simeq D_6$.
There are a number of ways of doing this.
One way is to use the action of $W(G_2)$ on the complexified Cartan subalgebra of $G_2$ to determine expressions for $u$ and $v$ in terms of a basis of $\tL$ coordinates on the Cartan subalgebra; a simpler way is to recognize that there are two components of the fixed point locus of the $W(G_2)$ action on $\C^2$ and to tune $B$ so that the discriminant of $Q$ factorizes appropriately; a more complicated way is to directly compute the Galois group of $Q$ \cite{Awtrey:Gal6}.
In any case, the result is
\begin{align}\label{Q G2}
   Q=\tL^6+u\tL^4+\tfrac14 u^2\tL^2 +\tfrac1{54}v ,
\end{align}
whose discriminant is $\sim v^3 (v-u^3)^2$. 

The dimensionless version of $Q$ is
\begin{align}\label{Th G2}
   \Th &= \tl^6 + w \tl^4 + \tfrac14 w^2\tl^2 +\tfrac1{54} ,&
   w & \doteq u v^{-1/3},
\end{align}
and $\tl \doteq \tL v^{-1/6}$. 
$\Th(\tl)=0$ is a 6-sheeted cover of the $w$-plane branched over $w^3=1$ as well as $w=\infty$. 
Label the 6 sheets $\tl^\pm_a$ for $a\in \Z_3$.
Since $\Th$ is even in $\tl$, we have $\tl^-_a = - \tl^+_a$.
\eqref{Th G2} and \eqref{attolt} then gives three linearly independent solutions to the integrability equation,
\begin{align}\label{ta G2}
    \ta^u_a &=-\frac{\tl^+_a}{6\tl_a^2+w}, &
    \ta^v_a &= 
    \frac{\tl^+_a}{2} \,
    \frac{2\tl_a^2+w}{6\tl_a^2+w} .
\end{align}
Since these expressions are odd in the $\tl$, the $\tl_a^-$ solutions are just the negative of the $\tl^+_a$ solutions.

The deck transformations of the $\tl^\pm_a$ are determined by a tedious analytic continuation along small counterclockwise loops around the branch points to be
\begin{align}\label{tl G2}
    \tl^\pm_a &\mapsto \tl_a^\pm, &
    \tl^\pm_{a\pm1} &\mapsto \tl^\pm_{a\mp1} , &
    a &\in \Z_3 ,
\end{align}
whether around $w=\z^a$ or $w=\infty$ if in the latter case the base point $w_*$ is such that $\frac{2\pi}{3} a < \arg(w_*) < \frac{2\pi}{3}(a{+}1)$.
Here $\z$ is a primitive cube root of unity. 
Since all these deck transformations square to the identity, we see from \eqref{ta G2} that the $\ta^k_a$ share the same deck transformations. 


We now determine the $2\times6$ $C$ intertwiner matrix \eqref{intertwiner} between the deck and automorphism twist representations of $W(G_2)$.
Since the deck transformations \eqref{tl G2} are all $\Z_2$ elements, the corresponding $\AutS \simeq D_6$ elements should also generate $Z_2$'s.  
There are seven $\Z_2$ subgroups in $D_6$, generated respectively by 
\begin{align}\label{AT G2}
  \pm A_n & \doteq \pm \bpm0&\z^n \\ \z^{-n}&0\epm, &
  n &\in \Z_3, &
  &\text{and}&
  &\bpm -1 & 0\\ 0 & -1\epm .
\end{align} 

The intertwiner condition \eqref{intertwiner}, for automorphism twist representation $A$ in the list of \eqref{AT G2} and permutation representation $D$ given by the deck transformations \eqref{tl G2}, yields, in the basis $(\tl^+_0, \tl^+_1, \tl^+_2, \tl^-_0, \tl^-_1, \tl^-_2)$, six distinct non-degenerate solutions up to overall normalization,
\begin{align}\label{C G2}
   C^n_\pm &= \bpm 
   \z & \z^2 &1 & -\z & -\z^2 & -1 \\ 
   \pm \z^{n} & \pm \z^{n+2} & \pm \z^{n+1} 
   & \mp \z^{n} & \mp \z^{n+2} & \mp \z^{n+1} \epm , & 
   n &\in \Z_3 .
\end{align}
Different choices of $n$ only amount to permuting the sheets as $\tl^\pm_a \to \tl^\pm_{a+n}$, so do not affect coefficients of the curve in the canonical frame, as they only depend meromorphically on $u$ and $v$. 
So, without loss of generality, one can take for example $C=C^1_\pm$. 
For the $+$ sign, the 1-form basis coefficient matrix in the automorphism frame is then computed to be
\begin{align}\label{form G2}
   \a^u_j &= 2\z \sum_{a\in\Z_3} (-\z^a)^{1+j} \,
   \frac{\tl_a^+}{6\tl_a^2+w} , &
   \a^v_j &= -\z\sum_{a\in\Z_3} (-\z^a)^{1+j} \, 
   \tl_a^+ \, \frac{2\tl_a^2+w}{6\tl_a^2+w} .
\end{align}
The solution with $C=C^1_-$ is found by reversing the signs of $\a^{u,v}_1$ relative to those of $\a^{u,v}_0$ in \eqref{form G2}.
Reverting to the CB scaling coordinates using \eqref{scaling}, the automorphism frame 1-form basis coefficient matrix is
\begin{align}\label{amat G2}
    a = \bpm 
    v^{-1/6}\a^u_0 & v^{-1/6}\a^u_1 \\ 
    v^{-5/6}\a^v_0 & v^{-5/6}\a^v_1  
    \epm .
\end{align}

Finally, it follows from \eqref{amap} and \eqref{AdetGG} that this automorphism frame solution is transformed to a canonical frame curve by acting on the automorphism frame curve \eqref{S G2} with the $\GL(2,\C)/\Z_3$ reparameterization given by $G= a^{-1} (\det a)^{1/3}$.
A computer-aided calculation using the $C^1_+$ solution \eqref{form G2} results in the canonical frame curve
\begin{align}\label{curve G2}
   y^2 =  \frac{1}{v(u^3-v)} \Bigl\{ 
   &\ \ \ \ x^6 \cdot 
   \ \ \ \ \ \bigl[ (c-2) u^3 +4 v \bigr]  
   \nn\\
   &+ x^5 \cdot 3^2 2 
   \ \ (c+2) u^2 v 
   \nn\\
   &+ x^4 \cdot 3^3 
   \ \ \ \bigl[ (c+10) u^3 +4c v \bigr] u v 
   \nn\\
   &+  x^3 \cdot 3^3 2^2  
   \bigl[ (3c+10) u^3 +2c v \bigr] v^2 
   \nn\\ 
   &+ x^2 \cdot 3^5 
   \ \ \ \bigl[ (c-10) u^3 + 4(c+5) v \bigr] u^2v^2 
   \nn\\
   &+ x \,\cdot\, 3^6 2 
   \ \, \bigl[(c-6) u^3 + 8 v \bigr] u v^3 
   \nn\\ 
   &+ \ \ \ \ \ 3^6 
   \ \ \ \bigl[ (c+2)u^6 -16 u^3 v +16 v^2 \bigr] v^3 
   \Bigr\} .
\end{align}
It satisfies the canonical frame integrability equation \eqref{inteq}, though in an intricate way. 

The $C^1_-$ choice of solution produces the same canonical frame curve except for the replacement $c \to -c$. 
Both solutions have discriminant $\propto (c^2-4)^3 (u^3 - v)^5 v^5$,
and the $c \to \pm 2$ curve degenerations are weak coupling limits.
We will see in section \ref{sec6} that, though not apparent in the canonical frame curve \eqref{curve G2}, the map $\s: c \mapsto -c$ is a symmetry of the theory, so the two solutions are equivalent.

\subsection{Non-split $BC_2$ curve}

The Weyl group is $W(BC_2) \simeq D_4$, so by \eqref{table} the corresponding automorphism frame curve is 
\begin{align}\label{S BC2}
   S(c): \qquad y^2=(x^2-1)(x^4+c x^2+1) ,
\end{align}
whose automorphism group is $\AutS=D_4$.
The $BC_2$ \Nf\ theory has CB scaling dimensions $\D_u=2$ and $\D_v=4$, so has coprime pair $(p,q)=(1,2)$ with $R=2$. 
The lowest-order scaling-homogeneous polynomial is
\begin{align}\label{Q BC2}
    Q &= \tL^4 + \sqrt2 u \tL^2 + \tfrac12 v,
\end{align}
where we have used the freedom to rescale $u$ and $v$ to choose the coefficients so that its discriminant is $\propto v(u^2-v)^2$.
As the discriminant locus has the two components expected from the $W(BC_2)$ orbifold action, $\GalQ = W(BC_2) \simeq D_4$, as can also be checked by direct computation of the Galois group \cite{Conrad:Gal34}.

The dimensionless version of $Q$ is
\begin{align}\label{Th BC2}
   \Th &= \tl^4+\sqrt2 w \tl^2 + \tfrac{1}{2} , &
   w &\doteq u v^{-1/2} ,
\end{align}
and $\tl \doteq \tL v^{-1/4}$.
$\Th(\tl)=0$ is a 4-sheeted cover of the $w$-plane branched over $w^2=1$ and $w=\infty$. 
Label the 4 sheets $\tl^\pm_a$ for $a\in\Z_2$.
Since $\Th$ is even in $\tl$, we have $\tl^-_a = - \tl^+_a$.
\eqref{Th BC2} and \eqref{attolt} then give
\begin{align}\label{ta BC2}
    \ta^u_a &= -\frac{\tl^+_a}2 \,
    \frac{1}{\sqrt2\tl_a^2+w} , &
    \ta^v_a &= \frac{\tl^+_a}4 \, 
    \frac{\sqrt2\tl_a^2+2w}{\sqrt2\tl_a^2+w} .
\end{align}
Since these expressions are odd in the $\tl$, the $\tl_a^-$ solutions are just the negative of the $\tl^+_a$ solutions.

The deck transformations of the $\tl^\pm_a$  along small counterclockwise loops around the branch points are
\begin{align}\label{tl BC2}
    \tl_1^\pm &\leftrightarrow \tl_2^\pm & 
    &\text{around $w=1$}, \nn\\
    \tl_1^\pm &\leftrightarrow \tl_2^\mp &
    &\text{around $w=-1$}, \\ 
    \tl_{1,2}^+ &\leftrightarrow \tl_{1,2}^- & 
    &\text{around $w=\infty$}. \nn
\end{align}
Since all these deck transformations square to the identity, it follows from \eqref{ta BC2} that the $\ta^k_a$ share the same deck transformations. 

Since the deck transformations \eqref{tl BC2} are all $\Z_2$ elements, the corresponding $\AutS \simeq D_4$ elements that the $2\times4$ $C$ matrix \eqref{intertwiner} intertwines should also generate $Z_2$'s.  
There are five $\Z_2$ subgroups in $D_4$, generated by 
\begin{align}\label{AT BC2}
   &\bpm0&i^n\\(-i)^n&0\epm , &
   n &\in \Z_4, &
   &\text{and} &
   &\bpm-1&0\\0&-1\epm .
\end{align} 
The intertwiner condition \eqref{intertwiner}, for automorphism twist representation $A$ in the list of \eqref{AT BC2} and permutation representation $D$ given by the deck transformations \eqref{tl BC2}, yields, in the basis $(\tl^+_0, \tl^+_1, \tl^-_0, \tl^-_1)$, a 2-parameter family of distinct solutions,
\begin{align}\label{C BC2}
   C(g,h) = \bpm g & h & -g & -h\\ 
   i h & i g & -ih & -ig\epm  .
\end{align}
For non-degeneracy the two parameters must satisfy $g\neq h$.
The overall normalization of $C(g,h)$ is arbitrary, so one of $g$ or $h$ can be set to 1 if it is non-zero.
Furthermore, the solutions with $g$ and $h$ exchanged are equivalent, since they just correspond to relabelling the $\tl^\pm_a$ sheets, thus $C(g,h) \simeq C(h,g)$.

But upon using $C(g,h)$ in \eqref{intertwiner} along with \eqref{ta BC2} to deduce the 1-form coefficient matrix,
\begin{align}\label{amat BC2}
    a = \bpm 
    v^{-1/4}\a^u_0 & v^{-1/4}\a^u_1 \\ 
    v^{-3/4}\a^v_0 & v^{-3/4}\a^v_1  
    \epm ,
\end{align}
and then using an $a^{-1} (\det a)^{1/3} \in \GL(2,\C)/\Z_3$ reparameterization, results in a canonical frame curve which is not holomorphic over the CB unless $gh=0$.
In particular, if $gh\neq0$ the canonical frame curve has $\sqrt v$ branch cuts; we explain why this is the case in the next section.

Therefore, there is a unique well-defined solution for the \Nf\ non-split $BC_2$ curve and 1-form given by the $C(1,0) \simeq C(0,1)$ intertwiner solution.
It results in the canonical frame curve
\begin{align}\label{curve BC2}
   y^2 = \frac{1}{v(v-u^2)} \Bigl\{ 
   & \ \ \ x^6 \cdot 
   \ \ \ \, \bigl[(c+2) v - 4 u^2 \bigr] u
   \nn \\
   &+x^5 \cdot 
   2^2 \bigl[(c+2) v +2(c-4) u^2\bigr] v
   \nn \\
   &+x^4 \cdot 
   2^2 \bigl[(11 c- 26) v +4(c-1) u^2\bigr] u v
   \nn \\
   &+x^3 \cdot 
   2^5 \bigl[(c-6) v +4(c-1) u^2\bigr] v^2
   \nn \\ 
   &+x^2 \cdot 
   2^4 \bigl[(11 c -26) v +4(c-1) u^2 \bigr] u v^2
   \nn \\
   &+x \,\cdot\, 
   2^6 \bigl[(c+2) v +2(c-4) u^2\bigr] v^3
   \nn \\ 
   &+ \ \ \ \ \ 2^6 \bigl[(c+2) v -4 u^2 \bigr] u v^3 \Bigr\} .
\end{align}
It has discriminant $\propto (c-2)^2(c+2)^6(u^2 - v)^5 v^5$.
The degenerations at $c=2$ and $c=\infty$ correspond to weak coupling limits.  The $c=-2$ degeneration does not correspond to a CB singularity, but rather to its geometry being split.

\section{Independence on the realization of the Galois group} \label{sec5}

In the last section we found a unique solution for the $A_2$ curve, two apparently inequivalent solutions for the $G_2$ curve, and a 1-parameter family of inequivalent solutions for the $BC_2$ curve.
But for $BC_2$, all but one of the curves turned out to be multi-valued over the CB, and so unphysical.
In this section we explain how to count from first principles the number of distinct curve solutions that are constructed using automorphism twists, and also the reason for the appearance of unphysical multi-valued curves in the way we implemented the automorphism twist in the last section.

Let us review the ingredients of the automorphism twist construction used in section \ref{sec2}. 
The curve was constructed from two objects: a fixed genus-2 curve, $S(c)$, and a polynomial $Q$. 
These have associated groups, the automorphisms of $S(c)$, $\AutS$, and the Galois group of $Q$, $\GalQ$, along with faithful representations, the embedding $\AutS \subset \GL(2,\C)$ specified by the chosen binary-sextic form of $S(c)$, and $\GalQ \subset \cS_{|Q|}$ given by the deck transformations permuting the sheets of $Q=0$. 
($|Q|$ is the degree of $Q$.)
Call these representations $G(\AutS)$ and $D(\GalQ)$, respectively.  
They have dimensions 2 and $|Q|$, respectively, and are typically reducible.

The automorphism frame orbifold construction proceeds by twisting the curve over the CB by elements of the automorphism twist group $\AT \subset \AutS$ \eqref{AT def} which correspond to the elements of the orbifold group fixing various codimension-1 subvarieties of the CB.  
Thus as an abstract group $\AT$ is the orbifold group.
We choose $Q$ by demanding that $\AT$ is isomorphic to $\GalQ$, so that the elements of the orbifold group are identified with the deck transformations around the branch points of the Riemann surface $Q=0$; the orbifold fixed point subvariety is then the discriminant locus of $Q$.

The restriction of $G(\AutS)$ to $G(\AT)$ furnishes a 2-dimensional representation of $\AT$, and thus there is an intertwiner $C$ as in \eqref{intertwiner}, a.k.a.\ an $\AT$-equivariant map, $C: D(\GalQ) \to G(\AT)$. 
To each $C$ corresponds a SW curve and one-form satisfying the SK integrability conditions.

We can count the distinct solutions for the intertwiner $C$ by decomposing $G(\AT)$ and $D(\GalQ)$ into irreducible representations as
\begin{align}
G(\AT) &\simeq \bigoplus_{j=1} g_j {\bf R_j}, &
D(\GalQ) &\simeq \bigoplus_{j=1} d_j {\bf R_j},
\end{align}
where $\bf R_i$ denote inequivalent irreducible representations of $\AT \simeq \GalQ$, and $g_j$ and $d_j$ denote the multiplicities with which they appear in $G(\AT)$ and $D(\GalQ)$.  
Define $|j| \doteq \dim({\bf R_j})$, one has the multiplicity $\sum_j g_j|j|=2$ and $\sum_j d_j|j|=|Q|$.  
By Schur's lemma, the matrix $C$ can be decomposed in terms of possibly nontrivial blocks of dimension $|i|\times |j|$ in a suitable basis, which is proportional to identity when ${\bf R_i} = {\bf R_j}$, and zero otherwise. 
Thus the number of independent constants contained in the $C$-matrix is $\sum_j d_j g_j - 1$, where the $-1$ accounts for the freedom to adjust the overall normalization of $C$.

For example, consider the non-split $BC_2$ curve.
There we have $\AT=\AutS=W(BC_2)=D_4$, which has a 2-dimensional irreducible representation, ${\bf R_0}$, and four 1-dimensional irreducible representations, ${\bf R_j}$, for $j\in\{1,2,3,4\}$. 
$G(D_4)$ is equivalent to ${\bf R_0}$, while a character computation shows that $D(D_4) = {\bf R_1} \oplus {\bf R_3} \oplus {\bf R_0}$. 
Thus only one constant appears in the intertwiner matrix $C$ which can be normalized to 1 without loss of generality.
Thus we expect there should be a unique solution to the intertwiner condition, and thus for the curve.

But we saw in section \ref{sec4} that the intertwiner for non-split $BC_2$ contains two undetermined constants (minus one overall normalization): there seems to be a mismatch. 
The source of this mismatch is that we computed $C$ not by using the polynomial $Q$ \eqref{Q BC2}, but rather using the rescaled polynomial $\Th$ \eqref{Th BC2}, which we recall here:
\begin{align}\label{QTh BC2}
    Q &= \tL^4 + \sqrt2 u \tL^2 + \tfrac12 v, \nn\\
    \Th &= \tl^4 + \sqrt2 w \tl^2 + \tfrac12, &
    w &\doteq u v^{-1/2} ,&
    \tl &\doteq v^{-1/4} \tL.
\end{align}
The point is that the Galois group of $Q$ (as an extension of the field of rational functions of $u$ and $v$) is not the same as the Galois group of $\Th$ (as an extension of the field of rational functions of $w$).
This is simply because the definition of $w$ involves the square root of $v$ which is not rational in $v$.
So while the Galois group of $Q$ is $\GalQ=D_4$, the Galois group of $\Th$ is the subgroup ${\rm Gal}\Th = \Z_2 \times \Z_2 \subset D_4$.
With respect to this subgroup the irreducible representations of $D_4$ decompose as
\begin{align}
    {\bf R_0}|_{\Z_2\times\Z_2} &= {\bf\til R_3} \oplus {\bf\til R_4} \nn\\
    {\bf R_1}|_{\Z_2\times\Z_2} = {\bf R_3}|_{\Z_2\times\Z_2} &= {\bf\til R_1} \nn\\
    {\bf R_2}|_{\Z_2\times\Z_2} = {\bf R_4}|_{\Z_2\times\Z_2} &= {\bf\til R_2}, 
\end{align}
where ${\bf\til R_j}$ for $j\in\{1,2,3,4\}$ are the four inequivalent 1-dimensional representations of $\Z_2\times\Z_2$.
Consequently, since $C$ now intertwines ${\bf\til R_3} \oplus {\bf\til R_4}$ with $2\cdot{\bf\til R_1} \oplus{\bf\til R_3} \oplus {\bf\til R_4}$,
there are two undetermined constants contained in $C$ (up to an arbitrary overall normalization), explaining the two-parameter family of intertwiners, $C(g,h)$, found in the last section.
But these solutions are designed to give canonical frame curves rational only in $w$, not in $u$ and $v$, and, indeed, for general values of $g$ and $h$ give curves containing square roots of $v$.
Only for $gh=0$ are the solutions rational in $u$ and $v$, giving the unique solution with intertwiner $C(1,0) \simeq C(0,1)$ (where we have used the freedom to choose the irrelevant overall normalization of the intertwiner).

In fact, this reduction, ${\rm Gal}\Th \subsetneq \GalQ$, also occurs for our choice \eqref{Th G2} of dimensionless CB parameter $w \doteq u v^{-1/3}$ in our calculation of the $G_2$ curve, though in this case this reduction happens not to result in extra-parameter families of non-rational curve solutions.

One can check that the same solutions for curves can be reproduced for other choices, $\til w$, of dimensionless CB parameter.
A natural candidate is to use 
\begin{align}
    \til w \doteq vu^{-q/p} ,
\end{align}
in terms of which the dimensionless polynomials become
\begin{align}
   G_2:\qquad\quad \til\Th &= \tl^6+\tl^4+\frac{\tl^2}{4}+\frac{\til w}{54} , \nn \\
   BC_2:\qquad\quad \til\Th &= \tl^4+\sqrt2\tl^2+ \frac{\til w}{2} .
\end{align}
With this choice one checks that the Galois groups are $D_6$ and $D_4$, respectively, i.e., the same as the Galois groups of the $Q$ polynomials. 
For example, in the $BC_2$ case, $\til\Th=0$ gives a 4-sheeted cover of the $\til w$ plane branched over the two points $\til w=0$ and $\til w=1$ with deck transformations
\begin{align}
    \tl_1^+ &\leftrightarrow \tl_1^- & 
    &\text{around $\til w=0$}, \\
    \tl_1^\pm &\leftrightarrow \tl_2^\pm & 
    &\text{around $\til w=1$}. \nn
\end{align}
As a result, there is only one inequivalent solution to the intertwiner, which is $C(1,0) \simeq C(0,1)$ (up to normalization), which is exactly the rational curve solution found in the last section.

These considerations do not explain the two apparently inequivalent solutions for the $G_2$ curve that we found.
We will see in the next section, in passing, that these two solutions are in fact equivalent, and their seeming inequivalence resulted from using a double covering of the moduli set of genus 2 curves with automorphism group $D_6$.

\section{Conformal manifolds and S-duality groups}
\label{sec6}

The conformal manifolds of \Nf\ theories with simple gauge algebras are 1-dimensional complex analytic spaces with orbifold singularities and punctures.
The S-duality groups of these theories are fundamental groups of these spaces in the orbifold sense \cite{Witten:1997sc}.
A familiar example of this is a theory with S-duality group $\SL(2,\Z)$, such as the $A_3$ \Nf\ sYM theory with global structure $(\SU(4)/\Z_2)_+$ (which forms a single S-duality orbit) \cite{Aharony:2013hda}.
In this case we identify the conformal manifold with a fundamental domain of the $\PSL(2,\Z)$ M\"obius action on the $\t$ complex upper half plane generated by $S: \t \to -1/\t$ and $T: \t \to \t+1$.
Any such a domain together with the identification of its boundaries by elements of $\PSL(2,\Z)$ has the topology of a 2-sphere with one puncture, one $\Z_2$ orbifold point, and one $\Z_3$ orbifold point.
For instance, in the ``usual'' fundamental domain $\cF \doteq \{|\Re\t| \le \frac12 \ \text{and}\ |\t|\ge1 \}$ with boundaries identified by $\PSL(2,\Z)$ transformations, the puncture is the $\t\to+i\infty$ limit, the $\Z_2$ point is the $\t=i$ fixed point of the $S$ element, and the $\Z_3$ point is the $\t = e^{i\pi/3}$ fixed point of the $TS$ element.
This domain is the conformal manifold of the theory, and its orbifold fundamental group is generated by the homotopy equivalence classes of paths, $\g_S$ and $\g_{TS}$, looping around the $\Z_2$ and $\Z_3$ orbifold points, respectively.
Thus the orbifold fundamental group has a presentation $\pi_1(\cF) = \langle\ A,B \  |\  A^2 = B^3=1 \ \rangle=\Z_2\ast \Z_3$, i.e., a free product of cyclic groups. 
One can see that this is isomorphic to $\PSL(2,\Z)$, thus the S-duality group $\pi_1(\cF) \simeq \PSL(2,\Z)$ is extracted from the (orbifold) topology of the conformal manifold.
Note that the topology of the conformal manifold cannot detect the center of the S-duality group in $\SL(2,\Z)$, which is charge conjugation in the field theory; however, the analysis from the CB geometry, given below, will be sensitive to this center.

We will apply this logic to compare the structure of the conformal manifolds of the $A_2$, $G_2$, and non-split $BC_2$,  \Nf\ sYM theories as predicted by the curves constructed in section \ref{sec4} to the predictions of their S-duality groups from \cite{Aharony:2013hda} and \cite{Dorey:1996hx, Argyres:2006qr}. 
Along the way we will also explain the occurrence of the two versions of the $G_2$ curve found in section \ref{sec4} which differed only by a change of sign of their coupling parameter $c$. 

\subsection{S-duality of rank 2 \Nf\ sYM theories}

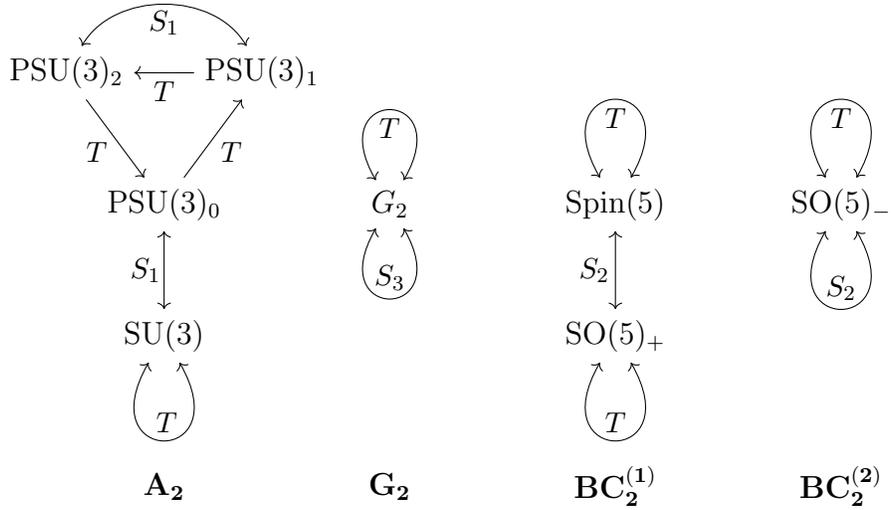
\begin{figure}
\begin{center}
\begin{tikzpicture}[scale=1]
\begin{scope}[scale=1, xshift=0cm]
\node (PSU0) at (0,1.75) {$\PSU(3)_0$};
\node (PSU1)at (1.3,3.5) {$\PSU(3)_1$};
\node (PSU2) at (-1.3,3.5) {$\PSU(3)_2$};
\node (SU) at (0,0) {$\SU(3)$};
\draw[<->] (SU) to node [xshift=-7pt, yshift=0pt, font=\small] {$S_1$} (PSU0);
\draw[<->] (PSU1) to [bend right=60] node [xshift=0pt, yshift=-7pt, font=\small] {$S_1$} (PSU2);
\draw[<->] (SU) to[in=-120, out=-60, looseness=10] node[xshift=1pt, yshift=7pt, font=\small]{$T$} (SU);
\draw[->] (PSU0) to node[xshift=7pt, yshift=-5pt, font=\small]{$T$} (PSU1);
\draw[->] (PSU1) to node[xshift=0pt, yshift=-7pt, font=\small]{$T$} (PSU2);
\draw[->] (PSU2) to node[xshift=-7pt, yshift=-5pt, font=\small]{$T$} (PSU0);
\node at (0,-2) {$\bf A_2$};
\end{scope}
\begin{scope}[scale=1, xshift=3cm]
\node (G2) at (0,1.75) {$G_2$};
\draw[<->] (G2) to[in=-120, out=-60, looseness=10] node[xshift=0pt, yshift=8pt, font=\small]{$S_3$} (G2);
\draw[<->] (G2) to[in=120, out=60, looseness=10] node[xshift=0pt, yshift=-7pt, font=\small]{$T$} (G2);
\node at (0,-2) {$\bf G_2$};
\end{scope}
\begin{scope}[scale=1, xshift=6cm]
\node (Spin) at (0,1.75) {$\Spin(5)$};
\node (SO+) at (0,0) {$\SO(5)_+$};
\draw[<->] (SO+) to node [xshift=-8pt, yshift=0pt, font=\small] {$S_2$} (Spin);
\draw[<->] (SO+) to[in=-120, out=-60, looseness=10] node[xshift=0pt, yshift=7pt, font=\small]{$T$} (SO+);
\draw[<->] (Spin) to[in=120, out=60, looseness=10] node[xshift=0pt, yshift=-7pt, font=\small]{$T$} (Spin);
\node at (0,-2) {$\bf BC_2^{(1)}$};
\end{scope}
\begin{scope}[scale=1, xshift=9cm]
\node (SO-) at (0,1.75) {$\SO(5)_-$};
\draw[<->] (SO-) to[in=-120, out=-60, looseness=10] node[xshift=0pt, yshift=8pt, font=\small]{$S_2$} (SO-);
\draw[<->] (SO-) to[in=120, out=60, looseness=10] node[xshift=0pt, yshift=-7pt, font=\small]{$T$} (SO-);
\node at (0,-2) {$\bf BC_2^{(2)}$};
\end{scope}
\end{tikzpicture}
\end{center}
\caption{S-duality orbits of rank 2 \Nf\ theories.}
\label{fig1}
\end{figure}

We start by reviewing (and, in the case of the $BC_2$ theories, refining) the predictions from \cite{Aharony:2013hda, Dorey:1996hx, Argyres:2006qr} for the S-duality groups of the rank 2 \Nf\ theories.
The four global structures of the $A_2$ \Nf\ sYM theory form a single S-duality orbit; there is a single global structure (and thus S-duality orbit) for the $G_2$ theory, and there are 3 global structures of the $BC_2\simeq\so(5)\simeq\sp(4)$ theory which fall into two orbits, which we call the $BC_2^{(1)}$ theory which is an orbit of two global structures, and the $BC_2^{(2)}$ theory which is an orbit of a single global structure. 
These orbits are shown in figure \ref{fig1} where we use the notation $\PSU(3) \doteq \SU(3)/\Z_3$.

From this orbit structure one can read off the S-duality group of each theory as a subgroup of the level-$q$ Hecke groups $H_{q} \doteq \langle S_\ell, T \, |\, S_\ell^2=(S_\ell T)^{q}=1 \rangle$.  
Their generators can be presented as $\PSL(2,\R)$ M\"obius transformations on the complex upper half $\t$ plane, $\cH_1$, by
\begin{align}\label{H2q}
    S_\ell : \t \mapsto -\frac1{\ell\t}, \qquad
    T : \t \mapsto \t +1, \qquad
    \t\in \cH_1,
\end{align}
where $\ell$ is related to $q$ by $\ell = 2(\cos(2\pi/q)+1)$.
Hecke groups are defined for all integers $q\ge3$, but are arithmetic only for $q=3,4,6$, for which $\ell=1,2,3$, respectively.
$\ell$ is the lacing number for the gauge Lie algebra,
\begin{align}\label{lacing}
    \ell=1 \quad \text{for}\quad A_2, \qquad
    \ell=2 \quad \text{for}\quad BC_2, \qquad
    \text{and}\qquad
    \ell=3 \quad \text{for}\quad G_2.
\end{align}
This realization of the S-duality groups as subgroups of the Hecke groups follows from the analysis in \cite{Dorey:1996hx, Argyres:2006qr}.

\begin{figure}
\begin{center}
\begin{tikzpicture}[scale=0.8]
\begin{scope}[scale=2, xshift=0cm]
\fill[fill=black!05] (-1.5,2.5) -- (-1.5,.866) arc (120:0:1) arc (180:60:1) -- (1.5,2.5) -- cycle;
\draw[very thick] (-1.5,2.5) -- (-1.5,.866);
\draw[very thick] (-1.5,.866) arc (120:0:1);
\draw[very thick] (0,0) arc (180:60:1);
\draw[very thick] (1.5,.866) -- (1.5,2.5);
\draw[->, black!35] (0,0)--(0,2.5);
\draw[->] (0,0)--(-1.8,0);
\draw[->] (0,0)--(1.8,0);
\draw[dashed] (-1,0) arc (180:0:1);
\draw[dashed] (-.5,0)--(-.5,2.5);
\draw[dashed] (.5,0)--(.5,2.5);
\node at (1,-.22) {$1$};
\node at (.5,-.22) {$\frac12$};
\node at (0,-.22) {$0$};
\node at (-.5,-.22) {-$\frac12$};
\node at (-1,-.22) {-$1$};
\node at (0,1.75) {$\cF_1$};
\node at (1,1.75) {$T\cF_1$};
\node at (-1,1.75) {$T^{-1}\cF_1$};
\node at (0,0.75) {$S_1\cF_1$};
\node[wd] at (0,2.6) {};
\node[wd] at (0,0) {};
\node[rd] at (1.5,.866) {};
\node[red] at (1.5,.65) {$\Z_3$};
\node at (0,-.75) {$\bf A_2$};
\end{scope}
\begin{scope}[scale=5, xshift=1.8cm]
\fill[fill=black!05] (-.5,1.) -- (-.5,.29) arc (150:30:.577) -- (.5,1.) -- cycle;
\draw[very thick] (-.5,1.) -- (-.5,.29);
\draw[very thick] (-.5,.29) arc (150:30:.577);
\draw[very thick] (.5,.29) -- (.5,1.);
\draw[->, black!35] (0,0)--(0,1.);
\draw[->] (0,0)--(-.7,0);
\draw[->] (0,0)--(.7,0);
\draw[dashed] (-.5,0)--(-.5,1.);
\draw[dashed] (-.577,0) arc (180:0:.577);
\draw[dashed] (.5,0)--(.5,1.);
\node at (.6,-.1) {$\frac1{\sqrt3}$};
\node at (.49,-.1) {$\frac12$};
\node at (0,-.1) {$0$};
\node at (-.49,-.1) {-$\frac12$};
\node at (-.66,-.1) {-$\frac1{\sqrt3}$};
\node at (0,0.78) {$\cF_3$};
\node[wd] at (0,1.05) {};
\node[rd] at (.5,.29) {};
\node[rd] at (0,.577) {};
\node[red] at (.5,.21) {$\Z_6\green{{\to}\Z_3}$};
\node[red] at (0,.49) {$\Z_2\green{{\to}1}$};
\node at (0,-.3) {$\bf G_2$};
\end{scope}
\begin{scope}[scale=4, xshift=0, yshift=-2.2cm]
\fill[fill=black!05] (-.5,1.5) -- (-.5,.5) arc (90:0:.5) arc (180:90:.5) -- (.5,1.5) -- cycle;
\draw[very thick] (-.5,1.5) -- (-.5,.5);
\draw[very thick] (-.5,.5) arc (90:0:.5);
\draw[very thick] (0,0) arc (180:90:.5);
\draw[very thick] (.5,.5) -- (.5,1.5);
\draw[->, black!35] (0,0)--(0,1.5);
\draw[->] (0,0)--(-1.,0);
\draw[->] (0,0)--(1.,0);
\draw[dashed] (-.707,0) arc (180:0:.707);
\draw[dashed] (-.5,0)--(-.5,.5);
\draw[dashed] (.5,0)--(.5,.5);
\node at (.7,-.13) {$\frac1{\sqrt2}$};
\node at (.5,-.13) {$\frac12$};
\node at (0,-.13) {$0$};
\node at (-.5,-.13) {-$\frac12$};
\node at (-.75,-.13) {-$\frac1{\sqrt2}$};
\node at (0,1.1) {$\cF_2$};
\node at (0,0.5) {$S_2\cF_2$};
\node[wd] at (0,1.55) {};
\node[wd] at (0,0) {};
\node[rd] at (.5,.5) {};
\node[red] at (.4,.375) {$\Z_2$};
\node at (0,-.4) {$\bf BC_2^{(1)}$};
\end{scope}
\begin{scope}[scale=4, xshift=2.25cm, yshift=-2.2cm]
\fill[fill=black!05] (-.5,1.5) -- (-.5,.5) arc (135:45:.707) -- (.5,1.5) -- cycle;
\draw[very thick] (-.5,1.5) -- (-.5,.5);
\draw[very thick] (-.5,.5) arc (135:45:.707);
\draw[very thick] (.5,.5) -- (.5,1.5);
\draw[->, black!35] (0,0)--(0,1.5);
\draw[->] (0,0)--(-1.,0);
\draw[->] (0,0)--(1.,0);
\draw[dashed] (-.707,0) arc (180:0:.707);
\draw[dashed] (-.5,0)--(-.5,.5);
\draw[dashed] (.5,0)--(.5,.5);
\node at (.7,-.13) {$\frac1{\sqrt2}$};
\node at (.5,-.13) {$\frac12$};
\node at (0,-.13) {$0$};
\node at (-.5,-.13) {-$\frac12$};
\node at (-.75,-.13) {-$\frac1{\sqrt2}$};
\node at (0,1.1) {$\cF_2$};
\node[wd] at (0,1.55) {};
\node[rd] at (0,.707) {};
\node[rd] at (.5,.5) {};
\node[red] at (0,.575) {$\Z_2\green{{\to}1}$};
\node[red] at (.5,.4) {$\Z_4\green{{\to}\Z_2}$};
\node at (0,-.4) {$\bf BC_2^{(2)}$};
\end{scope}
\end{tikzpicture}
\end{center}
\caption{Conformal manifolds of the four rank 2 \Nf\ simple absolute gauge theories $A_2$, $G_2$, $BC_2^{(1)}$, and $BC_2^{(2)}$ are the shaded regions with boundaries identified by $H_{q}$ elements.  
The open red circles are the distinct weak coupling limits (punctures) and the labelled red dots are the distinct points of global symmetry enhancement (orbifold points).
The green arrows indicate how the enhanced symmetry is broken at generic points on the CB.}
\label{fig2}
\end{figure}
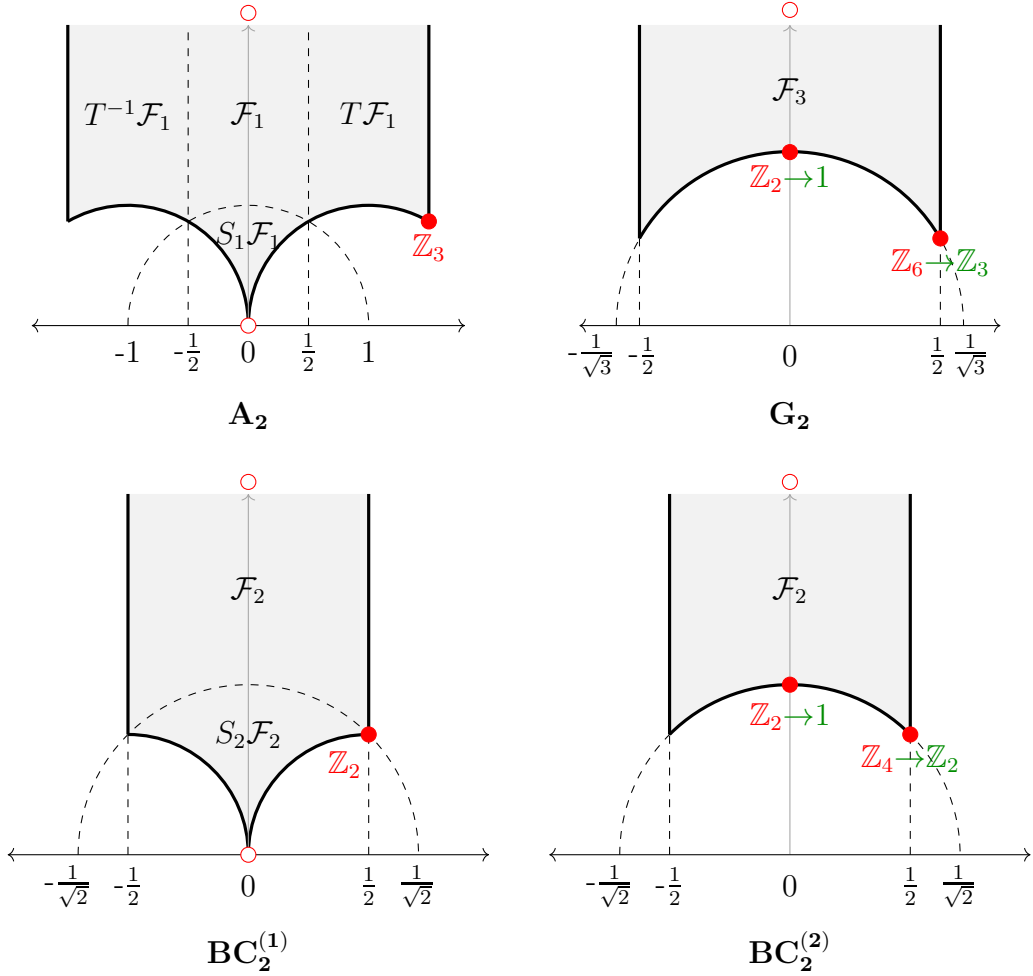

The resulting S-duality groups and conformal manifolds are shown in figure \ref{fig2}.
Here $\cF_\ell$ denotes a fundamental domain of the $H_q$ Hecke group action on $\cH_1$, and the shaded regions are a fundamental domain of the S-duality subgroup of $H_q$, in each case.
In words: 
\begin{itemize}
\item 
The $A_2$ S-duality group is the modular congruence subgroup $\Gamma^0(3) \subset \PSL(2,\Z) \simeq H_3$ generated by $T^3 :\t \to \t+3$ and $S_1TS_1:\t \to -1/\t$; 
its fundamental domain has 2 cusps at $\t\to i\infty$ and $\t\to0$ (the $\PSU(3)_{1,2,3}$ and $\SU(3)$ global structure weak coupling limits) and a single $\Z_3$ orbifold point at $\t=(3+i\sqrt3)/2$ fixed by $T^3S_1TS_1$.
\item
The $G_2$ S-duality group is all of $H_6$, generated by $T$ and $S_3$;
its fundamental domain has 1 cusp at $\t\to i\infty$, a $\Z_2$ orbifold point at $\t=i/\sqrt3$, and a $\Z_6$ orbifold point at $\t=(\sqrt3+i)/2\sqrt3$ fixed by $TS_3$.
\item
The $BC_2^{(1)}$ S-duality group is an index-2 congruence subgroup $H(4,2) \subset H_4$, generated by $T$ and $S_2TS_2$;
its fundamental domain has cusps at $\t\to i\infty$ and $\t\to0$, and a $\Z_2$ orbifold point at $\t=(1+i)/2$ fixed by $(TS_2)^2$.
\item
The $BC_2^{(2)}$ S-duality group is all of $H_4$, generated by $T$ and $S_2$;
its fundamental domain has 1 cusp at $\t\to i\infty$, a $\Z_2$ orbifold point at $\t=i/\sqrt2$, and a $\Z_4$ orbifold point at $\t=(1+i)/2$ fixed by $TS_2$.
\end{itemize}

\paragraph{Orbifold symmetries at special couplings.}

At a $\Z_n$ orbifold point on the conformal manifold with coordinate (coupling) $\t=\t_*$, the $\Z_n$ is a global symmetry of the field theory, which we will call the ``orbifold symmetry at coupling $\t_*$''.
Away from $\t_*$ this symmetry no longer exists:  it persists only as an identification of presentations of theories which one might have thought inequivalent.
In this sense, an orbifold symmetry at coupling $\t_*$ is gauged for $\t\neq \t_*$.
Thus there is a covering of the conformal manifold on which the orbifold symmetry acts freely except for a fixed point at $\t=\t_*$, and the conformal manifold is the orbifold of this covering space by the orbifold symmetry.

\subsection{Relation of the conformal manifold to the CB geometry}

How can we identify the conformal manifold from the low energy action on the CB?  
The low energy effective action on the CB is encoded in the SK geometry of the CB.
Define the ``moduli space of SK structures" to mean precisely the space of data given by a family of curves in automorphism frame together with a choice of 1-form basis $\{\Omega^u, \Omega^v\}$ as constructed in section \ref{sec4}, modulo equivalences which preserve the complex structure of the curve and the basis of 1-forms.
Then it is natural to guess that the moduli space of SK structures --- which is more or less the space of inequivalent CB effective actions --- coincides with the conformal manifold of the SCFT.
But the relation of the conformal manifold to the moduli space of the SK structures cannot be that straightforward.
There are a few reasons for this.

Firstly, the CB geometry encodes some SCFT observables, but there is no reason to suppose that it uniquely characterizes the SCFT.
Although we will not find any evidence of this at least in our rank 2 examples, there is the logical possibility that the moduli space of CB SK structures might be smaller than the conformal manifold --- i.e., the conformal manifold may be a covering of the SK structure moduli space.

Secondly, the orbifold symmetries of the SCFT at special points on its conformal manifold may be spontaneously broken on the CB.
In such a case, the moduli space of SK structures will be larger than the conformal manifold --- i.e., the SK structure moduli space will be a cover of the conformal manifold.
This does in fact occur in some of our examples. 

Indeed, it is known that $S_\ell$ S-duality generators for $\ell=2,3$ act not only on the conformal manifold as stated, but also act on the CB.
In particular, it was shown in \cite{Argyres:2006qr} that for $G_2$, the $S_3$ generator acts on the CB via
\begin{align}\label{brokenZ2}
    \varsigma : (u,\til v) \mapsto (u,-\til v)
\end{align}
where $u$ is the dimension-2 CB scaling coordinate, and $\til v$ is a certain%
\footnote{Note that dimension-6 coordinates are not uniquely (up to normalization) since $\til v+ \a u^3$ for arbitrary $\a\in\C$ is also a dimension-6 scaling coordinate.}
dimension-6 CB scaling coordinate.

A similar argument in the $BC_n$ case (almost identical to the one for $F_4$ given in \cite{Argyres:2006qr}) implies that for $BC_2$, the $S_2$ generator acts on the CB again via \eqref{brokenZ2}, but now with $u$ the dimension-2 CB scaling coordinate, and $\til v$ a certain dimension-4 CB scaling coordinate.%
\footnote{Similarly to the $G_2$ case, the $BC_2$ dimension-4 coordinate is not uniquely defined up to normalization.}

Since the orbifold symmetry generators at the orbifold points of the conformal manifolds of the $G_2$ and $BC_2^{(2)}$ theories (shown in figure \ref{fig2}) are all odd in $S_\ell$, a $\Z_2$ subgroup of each is spontaneously broken at generic points (those with $\til v \neq 0$) of their CBs, thus leaving the symmetries shown in green in figure \ref{fig2} as the unbroken orbifold subgroups.
Note that the generator of the $\Z_2$ orbifold symmetry at $t_*=(1+i)/2$ in the $BC_2$ theory is not spontaneously broken since its generator, $(TS_2)^2$, is even in $S_2$.

This means the moduli space of SK structures of the $G_2$ and $BC_2^{(2)}$ theories are double covers of their corresponding conformal manifolds.

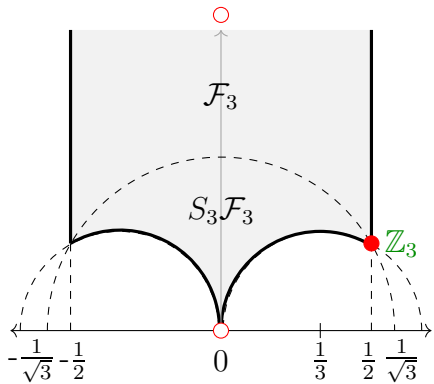
\begin{figure}
\begin{center}
\begin{tikzpicture}[scale=0.8]
\begin{scope}[scale=5, xshift=1.8cm]
\fill[fill=black!05] (-.5,1.) -- (-.5,.29) arc (120:0:.33) arc (180:60:.33) -- (.5,1.) -- cycle;
\draw[very thick] (-.5,.29) arc (120:0:.33);
\draw[very thick] (0,0) arc (180:60:.33);
\draw[very thick] (-.5,1.) -- (-.5,.29);
\draw[very thick] (.5,.29) -- (.5,1.);
\draw[->, black!35] (0,0)--(0,1.);
\draw[->] (0,0)--(-.7,0);
\draw[->] (0,0)--(.7,0);
\draw[dashed] (-.5,0)--(-.5,1.);
\draw[dashed] (-.667,0) arc (180:0:.33);
\draw[dashed] (-.577,0) arc (180:0:.577);
\draw[dashed] (.667,0) arc (0:180:.33);
\draw (.33,-.025)--(.33,.025);
\draw[dashed] (.5,0)--(.5,1.);
\node at (.6,-.1) {$\frac1{\sqrt3}$};
\node at (.49,-.1) {$\frac12$};
\node at (.33,-.1) {$\frac13$};
\node at (0,-.1) {$0$};
\node at (-.49,-.1) {-$\frac12$};
\node at (-.63,-.1) {-$\frac1{\sqrt3}$};
\node at (0,0.78) {$\cF_3$};
\node at (0,0.4) {$S_3\cF_3$};
\node[wd] at (0,1.05) {};
\node[rd] at (.5,.29) {};
\node[wd] at (0,0) {};
\node[red] at (.6,.29) {$\green{\Z_3}$};
\end{scope}
\end{tikzpicture}
\end{center}
\caption{Moduli space of the $G_2$ CB SK structures.}
\label{fig3}
\end{figure}

In the $G_2$ case, the double cover of $\cF_3$ is $\cF_3 \cup S_3\cF_3$, shown in figure \ref{fig3}.
It is the fundamental domain of the subgroup of $H_6$ generated by $T$ and $(S_3T)^2$.
It has two cusps and a $\Z_3$ orbifold point, just like the $A_2$ conformal manifold.
Indeed, it is not hard to see that this subgroup of $H_6$ consists of matrices of the form $\bspm a & b\\ 3c & d\espm \in \SL(2,\Z)$ which forms the $\Gamma_0(3)$ modular subgroup, which is isomorphic (via conjugation by $S_1$) to the $\Gamma^0(3)$ group of which the $A_2$ conformal manifold is the fundamental domain, shown in figure \ref{fig2}.
Thus we expect the moduli spaces of CB SK structures of the $A_2$ and $G_2$ theories to coincide.

In the $BC_2^{(1)}$ case, since the orbifold symmetry is not spontaneously broken on the CB, we expect the moduli space of its CB SK structures to coincide with the conformal manifold shown in figure \ref{fig2}.
This conformal manifold is the double cover of $\cF_2$ consisting of $\cF_2 \cup S_2\cF_2$, and has two cusps and a single $\Z_2$ orbifold point.
By contrast, in the $BC_2^{(2)}$ case a $\Z_2$ factor (generated by $S_2$) of the orbifold symmetries is spontaneously broken on the CB as shown in \ref{fig2}.
Thus the moduli space of the $BC_2^{(2)}$ CB SK structures is a double cover of the conformal manifold, so is given by $\cF_2 \cup S_2\cF_2$.
This coincides with the conformal manifold of the $BC_2^{(1)}$ theory.
Thus, we expect the moduli spaces of CB SK structures of the $BC_2^{(1)}$ and $BC_2^{(2)}$ theories to coincide as (orbifold) topological spaces.

Despite these coincidences of SK structure moduli spaces (between $A_2$ and $G_2$, and between $BC_2^{(1)}$ and $BC_2^{(2)}$), we will see that the SW curve and 1-form of these theories can distinguish between them because it detects the CB action of the broken symmetry generators.

\subsection{SK structure moduli space from the SW curve}

The SW curves derived in section \ref{sec4} come in 1-parameter families, with parameter $c\in \C \cup \{\infty\} \simeq \P^1$.
A point at $c=\infty$ is included simply because the curves are well-defined (unambiguous) in this limit.
Thus the moduli space of SK structures represented by these curves is covered by this $\P^1$.
Two or more different points $c \in \P^1$ could correspond to the same SK structure only if there is a $\GL(2,\C)/\Z_3$ reparameterization which fixes the form \eqref{table} of the curve but changes $c$.

In the case of the $A_2$ and $G_2$ automorphism frame curve, $y^2 = x^6 + c x^3 + 1$, a short computation shows there is one such reparameterization, $\s$, unique up to the $D_6$ automorphisms of the curve,
\begin{align}\label{sigma}
   &\s:&  (x,y) &\mapsto (-x,y) &
   (\w^0,\w^1) &\mapsto (\w^0,-\w^1) &
   c &\mapsto -c .
\end{align}
Thus the inequivalent complex structures of the curve are parameterized by $\P^1/\Z_2$ with the $\Z_2$ generated by $\sigma$, which has fixed points at $c=0$ and $c=\infty$.

If the curve were all that encoded the SK structure, then $\P^1/\Z_2$ would be the moduli space, and the points $c=0, \infty$ would be two $\Z_2$ orbifold points.
But the SK structure is specified by the SW curve and the SW 1-form.
As reviewed in section \ref{sec2}, the SW 1-form is equivalent to a choice of a basis, $\{\O^u,\O^v\}$, of holomorphic 1-forms on the curve (that varies over the CB).
But, by \eqref{repar}, all non-trivial $\GL(2,\C)/\Z_3$ reparameterizations act non-trivially on the canonical basis of 1-forms, $\{\w^0,\w^1\}$.
And so, since the $\{\O^j\}$ basis is non-degenerate, reparameterizations also act non-trivially on this basis as well.
Thus non-trivial $\GL(2,\C)/\Z_3$ reparameterizations which are not in the automorphism twist subgroup $\AT$ \eqref{AT def}, and thus $\s$ in \eqref{sigma} in particular, do not preserve the SK structure.
(The caveat concerning reparameterizations in $\AT$ will be explained below in the paragraph on cusps and orbifold points of the moduli space of SK structures.)

Thus, the moduli space of both the $A_2$ and $G_2$ SK structures is given by the $c\in\P^1$ without any identifications.
We will see in the $G_2$ case that $\s$ corresponds to the spontaneously broken $S_3$ S-duality generator, whereas in the $A_2$ case it does not correspond to any symmetry spontaneously broken on the CB.

In the case of the $BC_2$ automorphism frame curve, $y^2=(x^2-1)(x^4+c x^2+1)$, a short computation shows there are no reparameterizations which keep the form of the curve while changing $c$.
Thus, the moduli space of the $BC_2$ SK structures is also given by $c\in\P^1$ without any identifications.

\paragraph{Cusps and orbifold points in the moduli space of SK structures.}

We will now locate the cusps and orbifold points on these $c\in\P^1$ moduli spaces.

The cusps correspond to couplings at which charged states become massless at every point on the CB.
In field theory terms this corresponds to a gauge weak-coupling limit, while in terms of the CB geometry it corresponds to a degeneration of the curve (at all points on the CB) where the homology classes of the vanishing cycles are nontrivial.

The orbifold points correspond to couplings where there is an enhanced discrete global symmetry (again, at all points on the CB).  
Such a global symmetry is a symmetry of the SK structure, and so appears as a subgroup of the automorphism group of the curve which leaves the basis of 1-forms invariant.

By construction, all our curves have a large automorphism group --- $D_6$ in the case of the $A_2$ and $G_2$ curves, and $D_4$ in the case of the $BC_2$ curve. 
If these corresponded to global symmetries, they would imply discrete global symmetries at generic points on the conformal manifold.
As described in the paragraph below equation \eqref{sigma}, automorphisms which are $\GL(2,\C)/\Z_3$ reparameterizations act non-trivially on the $\{\omega^k\}$ basis of one-forms and so generically do not preserve the SK structure.  
But, by construction, automorphisms which are in the automorphism twist group $\AT$ \eqref{AT def} implement monodromies \eqref{amap} in the $a^j_k$ coefficient matrix --- relating the $\{\Omega^j\}$ basis which specifies the SK structure to the canonical $\{\omega^k\}$ 1-form basis --- precisely in order to compensate for the automorphism twist of the curve.
Thus automorphisms in $\AT$ preserve the SK structure.
However, they are not interpreted as global symmetries of the theory, but rather as gauge symmetries;  indeed, by construction they coincide with the Weyl group of the gauge Lie algebra.

So, denoting the group of automorphisms of the curve at $c$ by $\AutS_c$, it is only automorphisms in
\begin{align}\label{global H}
    H_c \doteq \AutS_c/\AT
\end{align}
which can have the interpretation as global symmetries of the theory at this value of $c$.%
\footnote{Strictly speaking, this is true only when $\AT$ is normal in $\AutS_c$.  The more general, correct, statement is given below in \eqref{normalizer}.}
An element $h\in H_c$ which preserves the $\{\Omega^j\}$ 1-form basis is interpreted as a global symmetry which is not spontaneously broken on the CB (i.e., has trivial action on the CB).
An element $h\in H_c$ that does not preserve the 1-form basis is interpreted as a global symmetry which is spontaneously broken on the CB if there is a global holomorphic scale-invariant reparameterization of the CB which compensates for the 1-form basis non-invariance.
Otherwise, $h$ cannot be interpreted as a global symmetry.

Global holomorphic scale-invariant reparameterizations of the CB are very tightly constrained.
Up to an overall irrelevant scaling, they can only be of the form
\begin{align}\label{CB reparam2}
    \varsigma: (u,v) \mapsto 
    (u, \a v + \b u^n),
\end{align}
for some complex coefficients $\a$, $\b$ and for $n$ a positive integer such that $n \D_u = \D_v$.
There is no such integer for $A_2$, thus $\b=0$; while for $G_2$ $n=3$ and for $BC_2$ $n=2$.
In order to possibly be a symmetry of the SK structure, $\varsigma$ must at least preserve the discriminant locus of the canonical frame curve.
The $A_2$ curve \eqref{curve A2} discriminant is $\propto (u^3-v^2)^5$,  
the $G_2$ curve \eqref{curve G2} discriminant is $\propto v^5(u^2-v)^5$, and the $BC_2$ curve \eqref{curve BC2} discriminant is $\propto v^5(u^3-v)^5$.
Thus the only possible broken symmetry actions on their CBs are
\begin{align}\label{CB reparam}
    \text{$A_2$ curve}&& 
    \varsigma:\qquad (u,v) &\mapsto 
    (u, -v), \nn\\
    \text{$G_2$ curve}&&
    \varsigma:\qquad (u,v) &\mapsto 
    (u, -v+u^3), \\
    \text{$BC_2$ curve}&&
    \varsigma:\qquad (u,v) &\mapsto 
    (u, -v+u^2). \nn
\end{align}
Note that these can be rewritten as CB maps of the form \eqref{brokenZ2} with the coordinate redefinitions $\til v \doteq v,\, v{-}u^3/2 ,\, v{-}u^2/2$ for the $A_2$, $G_2$, and $BC_2$ cases, respectively.

We will now analyze the moduli spaces of SK structures for the $A_2$, $G_2$, and $BC_2$ curves in turn, using their automorphism frame descriptions.

\paragraph{SK moduli space of the $A_2$ curve.}

The automorphism frame $A_2$ curve is given in \eqref{table}, and has automorphism group $\AutS=D_6$ for generic $c$, but has four special values, $c=0,\pm2,\infty$, where the automorphism group is enhanced and/or the curve degenerates.

The automorphism twist group is $\AT = \cS_3 \lhd \AutS$, so the group of possible global symmetries at a generic value of $c$ is $H = \AutS/\AT \simeq \Z_2$.
Indeed, $D_6 = \Z_2 \times \cS_3$, and the central $\Z_2$ is generated by the hyperelliptic automorphism which acts as 
\begin{align}\label{hyper aut}
    &\eta:&  (x,y) &\mapsto (x,-y) &
   (\w^0,\w^1) &\mapsto (-\w^0,-\w^1) &
   c &\mapsto c .
\end{align}
This therefore acts on the SW 1-form \eqref{sw1form} as $\L \mapsto -\L$, and so does not preserve the SK structure.

Therefore, if $\eta$ is a global symmetry, it must be broken on the CB, so must be compensated by the $A_2$ $\varsigma$ map \eqref{CB reparam}.
From \eqref{Q A2} it follows that $\varsigma: \til\L \mapsto -\til\L$, and so maps $w \to -w$ and $\til\l \to +\til\l$.
Since, by \eqref{form A2}, $\a^k_j$ are even in $w$, they are invariant under $\varsigma$, and so from \eqref{amat A2} we learn that $\varsigma: (a^u_i, a^v_i) \mapsto (-a^u_i, a^v_j)$.  
The definition of the 1-form basis $\Omega^k \doteq a^k_i \omega^i$ implies that the combined action of the hyperelliptic automorphism \eqref{hyper aut} and the CB action $\varsigma$ is $\eta\circ\varsigma: (\O^u,\O^v) \mapsto (\O^u,-\O^v)$.
This is precisely the action which preserves the SW 1-form, 
\begin{align}
    \eta\circ\varsigma:
    \L \doteq \D_u u\, \O^u+\D_v v\, \O^v
    \mapsto \D_u u\, \O^u+\D_v (-v)\, (-\O^v)
    = \L,
\end{align}
and thus the SK structure.

This shows that at generic values of the coupling $c$ the $A_2$ theory has a $\Z_2$ global symmetry which is spontaneously broken on the CB.
Physically this symmetry is just the charge conjugation symmetry of the $A_2$ theory, whose action on the CB was pointed out in \cite{Bourget:2018ond, Argyres:2018wxu}.

Note, by contrast, that the $\s$ curve reparametrization \eqref{sigma} (which gives the $c \to -c$ involution of the conformal manifold) does not act on the 1-form basis in the same way that the hyperelliptic involution $\eta$ does, so cannot be compensated by the $\varsigma$ CB involution.  
Thus there is no equivalence of the $A_2$ SK structure under $c\to-c$.

We now turn to the special points of the conformal manifold, $c=0,\pm2,\infty$ where the curve degenerates or its automorphism group enhances.

At $c=\pm2$ the curve degenerates by becoming a perfect square, implying that it has vanishing cycles which are homologically non-trivial.
This is the signature of a weak-coupling limit, since it implies the existence of additional massless charged degrees of freedom everywhere on the CB.
This gives the expected two weak-coupling cusps in the moduli space.

At $c=0$ the curve is not degenerate, but has an enhanced automorphism group $\AutS_0 \simeq \Z_3 \rtimes D_4$; see, e.g., Appendix B of \cite{Argyres:2022fwy}.
The $D_6$ automorphism group of the generic point must be a subgroup of $\AutS_0$, and indeed one finds%
\footnote{\label{ftnt GAP}We used GAP \cite{GAP4} and GroupNames \cite{GroupNames} for many of the finite group computations quoted in this section.} 
that it is a normal subgroup with $H_0 = \AutS_0 / D_6 \simeq \Z_2$, and furthermore that $\AutS_0$ is a split extension $\AutS_0 \simeq D_6 \rtimes \Z_2$ with the $\Z_2$ acting via the outer automorphism of $D_6$.  
This is precisely the $\s$ map \eqref{sigma}, which is now an automorphism since $c=0$ is a fixed point of $c\mapsto-c$.
But we have just seen that $\s$ acts non-trivially on the 1-form basis and can not be compensated by the CB involution $\varsigma$ \eqref{CB reparam}.
Therefore this automorphism group enhancement, $H_0$, does not correspond to a global symmetry, broken or not on the CB, and thus there is no orbifold point at $c=0$.

At $c=\infty$ the curve degenerates by developing a trivial homology vanishing cycle which pinches the genus-2 curve to a bouquet of two elliptic curves each with a $\Z_6$ automorphism group.%
\footnote{This is apparent from the form of the curve since, as $c\to \infty$, the $y=0$ roots of the curve are proportional to cube roots of unity, $x^3 \approx -c$ and $x^3 \approx -c^{-1}$.}
Thus this degeneration does not correspond to a weak-coupling cusp on the moduli space since although the genus-2 Riemann surface degenerates here, its Jacobian variety does not: it simply splits into the product of two $\Z_6$ tori.
The automorphism of this split rank-2 abelian variety is thus $\AutS_\infty = \Z_2 \ltimes (\Z_6 \times \Z_6)$. 
The $\Z_2$ factor acts via the automorphism that interchanges the two $\Z_6$ torus factors, so $\AutS_{\infty} = \Z_6 \wr \Z_2$, the wreath product of $\Z_6$ by $\Z_2$.
The generic $\AutS=D_6$ automorphism group is a subgroup of $\AutS_\infty$.
Indeed, it is a normal subgroup with $H_\infty \doteq \AutS_\infty /D_6 \simeq \Z_6$, again giving a split extension where the $\Z_6$ acts via a morphism $\f: \Z_6 \to \Z_2 = {\rm Out}(D_6)$.
In particular, the $\Z_2 ={\rm im}(\Z_6)$ acts via the outer automorphism of $D_6 = \AutS$ which we saw above has the $\s$ action \eqref{sigma} on the 1-form basis.
And as we showed above, this action cannot be compensated by a CB $\varsigma$ involution \eqref{CB reparam}.
Thus this $\Z_2\subset H_\infty$ is not a symmetry and does not contribute to an orbifold group at $c=\infty$.

The remaining $\Z_3 = {\rm ker}(\f) \subset H_\infty$ is central.  
Indeed $\AutS_\infty \simeq \Z_3 \times (\Z_3 \rtimes D_4) = \Z_3 \times \AutS_0$ is a direct product, and so this $\Z_3$ automorphism leaves the 1-form basis invariant.
Thus there is an enhanced $\Z_3$ global symmetry at $c=\infty$, and so this point is a $\Z_3$ orbifold point of the SK moduli space.

In summary, the SK moduli space is a $\P^1$ with two weak-coupling punctures, one $\Z_3$ orbifold point, and also has at every point a $\Z_2$ symmetry which is spontaneously broken on the CB.
This reproduces exactly the expected $A_2$ conformal manifold, shown in figure \ref{fig2}.

\paragraph{SK moduli space of the $G_2$ curve.}

Since the automorphism frame curve for the $G_2$ theory is the same as for the $A_2$ curve, it has the same weak-coupling cusps at $c=\pm2$ and automorphism enhancements at $c=0,\infty$.
But, because the SW 1-form for the $G_2$ curve is different, the interpretation of those automorphism enhancements as global symmetries is changed.

First, the automorphism twist group is the whole curve automorphism group, $\AT = \AutS \simeq D_6$, so the hyperelliptic automorphism $\eta$ \eqref{hyper aut} preserves the SK structure by construction.
Indeed, $\eta$ had the interpretation as charge conjugation in the $A_2$ theory, but charge conjugation is part of the gauge group in a $G_2$ gauge theory.
So there is no $\Z_2$ global symmetry at generic $c$.

Relatedly, the $G_2$ CB reparameterization $\varsigma$ \eqref{CB reparam} is \emph{not} a symmetry of the SK structure.
For although it preserves (by construction) the zeros of the canonical frame curve discriminant and their multiplicities, it does not preserve the 1-form basis.
The easiest way to see this is to note that the 1-form basis was constructed in automorphism frame in terms of the roots of the polynomial $Q$ in \eqref{Q G2} with Galois group $D_6$.
But while $\varsigma$ interchanges the $v=0$ and $v-u^3=0$ zeros of the discriminant of $Q$, these zeros have different multiplicities, and so $\varsigma$ is not a symmetry of $Q$ and so, a fortiori, not of the 1-form basis either.
Physically, the two components of the discriminant locus have distinct (weak-coupling) physical interpretations on the CB: one as the locus where an $\su(2) \subset G_2$ corresponding to a long root is unbroken, and the other where an $\su(2)$ factor corresponding to a short root is unbroken.

By contrast, the $\s$ curve reparameterization \eqref{sigma} which gives the $c\to-c$ involution of the conformal manifold, is now an equivalence of $G_2$ SK structures.
The reason is that since $\s$ is a reparameterization, it relates curves with the same complex structure.
And we found two distinct automorphism frame solutions for the $G_2$ curve, referred to as the $C^1_+$ and $C^1_-$ solutions in section \ref{sec4}.
As noted below \eqref{form G2} these solutions differ in automorphism frame by $(\a_0^{u,v}, \a_1^{u,v}) \to (\a_0^{u,v}, -\a_1^{u,v})$, which, combined with the $\s$ action on the canonical 1-form basis \eqref{sigma} shows that the SW 1-form is left invariant.
At generic $c$ it cannot be interpreted as a symmetry, since it changes the $c$ parameter.
But at the fixed points of $\s$, namely $c=0,\infty$, it shows that the two SK structures, $C^1_+$ and $C^1_-$, are equivalent.  
Since they are not identical, this equivalence must be realized via an involution of the CB, which, as argued earlier, can only be the $\varsigma$ involution \eqref{CB reparam}.   
(Note that the argument of the previous paragraph that $\varsigma$ is not a symmetry at generic $c$ does not apply at the $c=0,\infty$ points because at these points $\varsigma$ can relate the two distinct equivalent SK structures.)

The analysis of the automorphism group enhancements at $c=0$ and $\infty$ now immediately follows from the analysis in the $A_2$ case.
At $c=0$, $H_0 \doteq \AutS_0/\AutS = \Z_2$ which is the $\s$ automorphism which is a $\Z_2$ symmetry spontaneously broken on the CB.
At $c=\infty$, $H_\infty \doteq \AutS_\infty/\AutS = \Z_2 \times \Z_3$ where the $\Z_2$ factor is again generated by the $\s$ automorphism (a $\Z_2$ global symmetry spontaneously broken on the CB) and the $\Z_3$ is a global symmetry unbroken on the CB.

Thus the final picture of the $G_2$ SK moduli space is that of a $\P^1$ with two weak-coupling cusps and a single $\Z_3$ orbifold point, just as in the $A_2$ case (shown in figure \ref{fig2}) and as predicted for $G_2$ (as shown in figure \ref{fig3}).
But our analysis is sensitive enough to distinguish between the $A_2$ and $G_2$ cases since we also found in the $G_2$ case that at the $\Z_3$ point ($c=\infty$) there is a $\Z_2$ symmetry spontaneously broken on the CB, and at a regular point ($c=0$) there is another $\Z_2$ symmetry spontaneously broken on the CB, and furthermore, that these $\Z_2$ actions extend to a $\Z_2$ involution ($c\to-c$) of the SK moduli space.
This implies that the conformal manifold is the $c \sim -c$ quotient of the SK moduli space, giving a conformal manifold with a single weak-coupling cusp, a $\Z_6$ point and a $\Z_2$ point, as predicted by S-duality, and shown in figure \ref{fig2}.

\paragraph{SK moduli space of the $BC_2$ curve.}

The S-duality prediction for the orbifold topology of the moduli space of SK structures of the $BC_2^{(1)}$ and $BC_2^{(2)}$ theories, as reviewed in the previous subsection, is that they are both $\P^1$'s with two weak-coupling cusps and a single $\Z_2$ orbifold point, as shown for the $BC_2^{(1)}$ conformal manifold in figure \ref{fig2}.
They are distinguished by the $BC_2^{(2)}$ theory having a regular (non-orbifold) point with a $\Z_2$ symmetry which is broken on the CB, and that the symmetry group of the $\Z_2$ orbifold point is enhanced to $\Z_4$, a $\Z_2$ subgroup of which is spontaneously broken on the CB.
We will now show that our genus-2 (non-split) $BC_2$ curve realizes the $BC_2^{(1)}$ S-duality orbit and not the $BC_2^{(2)}$ one.

Recall that the $BC_2$ automorphism frame curve is $y^2 = (x^2-1)(x^4+cx^2+1)$.  
It has automorphism group $\AutS=D_4$ at generic values of $c$; $D_4$ is the $BC_2$ Weyl group.
Unlike the $A_2/G_2$ curve, there is no reparameterization of the curve preserving its form but acting non-trivially on $c$, so there is no analog of the $\s$ map \eqref{sigma}.
Also, the automorphism twist group is the whole of $\AutS$, so there is no possibility of an additional discrete global symmetry group (beyond the Weyl group, which is part of the gauge group) at generic values of $c$.
Relatedly, the only possible CB involution \eqref{CB reparam} cannot be a symmetry of the SK structure because, just as in the $G_2$ case, it does not preserve the discriminant of the $Q$ polynomial \eqref{Q BC2} used to construct the 1-form basis.

By inspection, the $BC_2$ curve has degenerations or enhanced automorphisms at special couplings $c=1$, $\pm2$, and $\infty$.
At $c=+2$ the curve degenerates to $y^2=(x^2-1)(x^2+1)^2$, which has vanishing cycles surrounding an even number of branch points, so are non-trivial in homology.
At $c=\infty$ the curve degenerates to $\til y^2=(x^2-1)x^2$ (after appropriately rescaling the $y$ coordinate), which has vanishing cycles at $x=0$ and $x=\infty$ which are also homologically non-trivial.
These are thus weak-coupling degenerations, corresponding to cusps at $c=2$ and $c=\infty$.

At $c=1$ the curve becomes $y^2=x^6-1$ which has enhanced automorphism group $\AutS_1 = \Z_3 \rtimes D_4$; see, e.g., \cite{Argyres:2022fwy}.
Since $\AutS_1$ is an enhancement of $\AutS \simeq D_4$ we must have $D_4 \subset \AutS_1$, which is easily checked to be true.
But, unlike at the analogous point in the $A_2/G_2$ moduli space, $\AutS$ is not a normal subgroup.
In this case the correct notion of what elements of the enhanced automorphism group at $c=*$ can be interpreted as possible global symmetries, $H_*$, is 
\begin{align}\label{normalizer}
    H_* = N_{\AutS_*}(\AT)/\AT ,
\end{align}
where $N_G(S)$ denotes the normalizer of $S$ in $G$.   
Indeed, the normalizer is the largest subgroup of $G$ for which $S$ is a normal subgroup.
Recall that we quotient by $\AT$ in \eqref{normalizer} because the automorphism twist elements are part of the gauge symmetry.

In our case of $\AT=\AutS=D_4$ and $\AutS_1=\Z_3 \rtimes D_4$, one finds that $N_{\Z_3 \rtimes D_4}(D_4) = D_4$, and so the group of possible global symmetries at $c=1$ is trivial.
Thus there is no orbifold point there and also no additional symmetry which might be spontaneously broken on the CB.

Finally, at $c=-2$ the curve degenerates to $y^2=(x^2-1)^3$ which has vanishing cycles at $x=\pm1$ surrounding an odd number of branch points, and so are homologically trivial.
This degeneration thus does not correspond to a singularity in the low energy effective action on the CB, but rather simply to the Jacobian of the genus-2 Riemann surface becoming split.
A short calculation keeping the subleading contributions to the curve in the $c\to-2$ limit shows that the Jacobian splits as the product of two tori each with automorphism group $\Z_4$ (i.e., with modulus $\t=i$), so the automorphism group there is $\AutS_{-2} = (\Z_4 \times \Z_4) \rtimes \Z_2$.
The $\Z_2$ factor acts via the automorphism that interchanges the two $\Z_4$ torus factors, so $\AutS_{-2} = \Z_4 \wr \Z_2$, the wreath product of $\Z_4$ by $\Z_2$.  

Now $D_4 \subset \Z_4 \wr \Z_2$ in two different ways, one as a normal subgroup and one not.
One easy way of working out which of these two $D_4$ subgroups corresponds to $\AutS$ is as follows.
The Jacobian variety of the $BC_2$ automorphism frame curve is a rank-2 principally polarized abelian variety.
Describe this variety by its modulus in the rank-2 Siegel space, i.e., as a symmetric complex matrix with positive imaginary part, $\t \doteq \bspm z_1 & z_3\\ z_3 & z_2 \espm$.
One parameterization of the $\t$ moduli of Jacobians with $\AutS=D_4$ is
\begin{align}\label{tau d4}
    \t = \frac12 \bpm z-z^{-1} & z+z^{-1} \\ z+z^{-1} & z-z^{-1} \epm .
\end{align}
This follows from the classification of \cite{gottschling1961fixpunkte} of fixed loci of the $\Sp(4,\Z)$ action on the rank 2 Seigel half-space, discussed in appendix C of \cite{Argyres:2022fwy}.
The parameter $z$ is some appropriate meromorphic function of $c$.
The automorphism group of this locus is given in \cite{gottschling1961fixpunktuntergruppen} as the subgroup of $\Sp(4,\Z)$ generated by
\begin{align}\label{sp4Z d4 gens}
   D_4 \simeq \AutS = \left\langle\ \bspm 0&1&0&0\\ 1&0&0&0\\ 
   0&0&0&1\\ 0&0&1&0 \espm \ ,\  
   \bspm 0&0&-1&0\\ 0&0&0&1\\ 1&0&0&0\\ 0&-1&0&0 \espm\ \right\rangle .
\end{align}
At $z=i$, $\t = \bspm i & 0 \\ 0 & i \espm$, so this corresponds to the point $c=-2$.
At this point the automorphism group of the abelian variety is \cite{gottschling1961fixpunktuntergruppen} 
\begin{align}\label{sp4Z aut gens}
   \Z_4 \wr \Z_2 \simeq \AutS_{-2} =\left\langle D_4 \ ,\  
   \bspm -1&0&0&0\\0&0&0&-1\\0&0&-1&0\\0&1&0&0 \espm\ \right\rangle .
\end{align}
A short computer-aided calculation\textsuperscript{\ref{ftnt GAP}} then shows that this $D_4$ subgroup is normal in $\AutS_{-2}$ and, furthermore, $\AutS_{-2}$ is a split extension of $D_4$ by the quotient.
So the group of possible global symmetries at $c=-2$ is
\begin{align}\label{BC2 Z4 symm}
    H_{-2} = \AutS_{-2} / \AutS = \Z_4 ,
\end{align}
with the $\Z_4$ factor acting on $\AutS = D_4$ via a morphism 
\begin{align}\label{H-2 morphism}
    \f: \Z_4/\Z_2 \to \Z_2 = {\rm Out}(D_4) .
\end{align}

We now want to discern what subgroup of $H_{-2} \simeq \Z_4$ acts trivially on the 1-form basis, and so is an unbroken global symmetry at $c=-2$.
A problem is that our curve description breaks down at $c=-2$, making it difficult to compute the $H_{-2}$ action on the 1-form basis.
A work-around (which we implicitly already used in the $A_2/G_2$ case in the analysis of the $H_\infty$ action) is to identify $\AutS \simeq D_4$ as a subgroup of $\GL(2,\C)/\Z_3$ via its action \eqref{AdetGG} on the 1-form basis away from $c=-2$, and then identify the $\Z_4$ with elements of $\GL(2,\C)/\Z_3$ via their action on $D_4$ by conjugation (i.e., automorphisms of $D_4$ as a subgroup of $\GL(2,\C)/\Z_3$).
The automorphism group of the $BC_2$ curve is computed in appendix B of \cite{Argyres:2022fwy} to be the subgroup of  $\GL(2,\C)/\Z_3$ generated by
\begin{align}\label{gl2 d4 gens}
    \AutS \simeq D_4 = \left\langle\ 
    \bpm 1&0\\ 0&-1 \epm \ ,\  
    \bpm 0&i\\ i&0 \epm\ \right\rangle .
\end{align}
Its outer automorphism group is then (up to conjugation by inner automorphisms)
\begin{align}\label{gl2 out gens}
    {\rm Out}(D_4) \simeq \Z_2 = \left\langle\ 
    \bpm 0&1\\ 1&0 \epm \ \right\rangle ,
\end{align}
whose generator thus acts on the canonical 1-form basis by $(\omega_1, \omega_2) \mapsto (\omega_2,\omega_1)$. 
${\rm Out}(D_4) = \f(H_{-2})$ therefore is not an unbroken symmetry of the SK structure at $c=-2$, and, since we argued above that there can be no broken symmetry action on the $BC_2$ CB, it is simply not a symmetry at all.
On the other hand, the kernel of the $\f$ morphism in \eqref{H-2 morphism}, ${\rm ker}(\f) = \Z_2$, has no action on $D_4$ and so it is in the centralizer of $D_4 \subset \GL(2,\Z)/\Z_3$, so its elements must be proportional to the identity matrix. 
The only such elements of order 2 are $\pm 1$ which act trivially on the 1-form basis and so also on the SK structure (since $-1 \in \AutS$).
Thus there is a $\Z_2$ global symmetry at $c=-2$, unbroken on the CB, and no further broken symmetry enhancement there.

In summary, this analysis precisely matches the S-duality prediction reviewed in the previous subsections for the $BC_2^{(1)}$ theory: two weak-coupling cusps, a single $\Z_2$ orbifold point, and no additional symmetries at any points of the SK structure moduli space which are spontaneously broken on the CB.
By contrast, the prediction for the $BC_2^{(2)}$ theory --- namely, the same weak-coupling and orbifold points but with a $\Z_4$ symmetry at the orbifold point spontaneously broken on the CB down to $\Z_2$, and an additional point with a $\Z_2$ symmetry completely broken on the CB --- is not realized.
We conclude that the (non-split) genus-2 $BC_2$ curve we have constructed describes the $BC_2^{(1)}$ S-duality orbit, and not the $BC_2^{(2)}$ orbit.

\section{Outlook, and comparison to previous \Nts\ curves}\label{sec7}

In conclusion, we have derived the genus 2 SW curves for two \Nf\ rank 2 sYMs, namely, the $G_2$ and the non-split $BC_2$ ones, using the automorphism twist approach. 
We have checked that these curves agree with S-duality predictions as long as the $BC_2$ curve is identified with the $BC_2^{(1)}$ theory.

These computations serve mainly as test cases for developing the automorphism twist approach with the eventual goal of applying these techniques to constructing more general isotrivial CB geometries. 
The isotrivial class includes all SCFTs with $\cN{\geq}3$ and many \Nt\ SCFTs as well \cite{Cecotti:2021ouq}. 
While for the \Nf\ isotrivial geometries discussed here, which are SK orbifolds without complex singularities, the automorphism approach may seem unnecessarily complicated, it nevertheless promises to be useful for non-orbifold isotrivial geometries and ones with complex singularities, which can occur for \Nt\ SCFTs.

In the rest of this section we compare the \Nf\  solutions found here to the previously-known \Nts\ $BC_2$ and $G_2$ solutions coming from integrable system constructions.  
In the massless limit these describe the CBs of \Nf\ theories.
We find that their constructions provide curves in the $BC_2$ case that only applies to the split theory, and gives a non-principal Dirac pairing in the $G_2$ case, similar to the $A_2$ case considered in \cite{Argyres:2022kon}.
Thus they fail to correspond to curves for either of the theories we have derived here.

CB geometries of \Nts\ theories with non-simply laced gauge algebras have been constructed from the symplectic geometry of the holomorphic phase spaces of twisted elliptic Calogero-Moser (CM) integrable systems \cite{DHoker:1997hut, DHoker:1998rfc, DHoker:1998xad, DHoker:1998zuv, Bordner:1998sw, Bordner:1998xs, Bordner:1998xsa}.  
Spectral curves associated to Lax pair presentations of these systems are interpreted as SW curves of the CB geometry, and the canonical 1-form (a.k.a., symplectic potential) for the symplectic structure on the phase space of the integrable system is interpreted as the SW 1-form.
In general the Lax presentation and associated spectral curve is not uniquely determined, and in the discussions of elliptic CM systems in this context depends on a choice of an orbit of a point in $\C^r$ under the complexified reflection representation action of the Weyl group of the Lie algebra.%
\footnote{This choice appears in \cite{DHoker:1997hut, DHoker:1998rfc, DHoker:1998xad, DHoker:1998zuv, Bordner:1998sw, Bordner:1998xs, Bordner:1998xsa} as a choice of faithful representation of the Lie algebra.
In what follows we recast this in terms of Weyl group orbits.
The relation between the two is that the weights of a representation give a finite union of Weyl orbits.
More generally, the construction of elliptic CM systems depends only on the Weyl group, and Lax presentations and spectral curves can also be determined without reference to Lie-theoretic data.
For example, elliptic CM systems can be generalized to ones based on arbitrary crystallographic complex reflection groups \cite{EFMV11ecm} for which no analog of Lie algebra representation exists, and these generalized elliptic CM systems describe CB geometries \cite{Argyres:2023tfx}.
}
Given such a choice, the resulting spectral curve has genus given by the size, $N$, of the orbit, which is larger than the rank, $r$, of the CB (which is the dimension of the Weyl group reflection representation).
So the spectral curve must be supplied with additional data in order for it to be interpreted as a SW curve for the CB geometry.

This extra data is given by a choice of a rank-$2r$ symplectic sublattice of the 1-homology lattice of the Jacobian variety of the spectral curve.
The 1-homology lattice of ${\rm Jac}(\S_N)$ is a rank-$2N$ symplectic lattice which inherits an action of the Weyl group from the construction of the spectral curve.
The rank-$2r$ sublattice is a Weyl-invariant sublattice of the 1-homology lattice.
The identification of these invariant sublattices does not seem to have been performed explicitly in the literature, so we will do this below.

For $w_i\in\C^r$, $i=1,\ldots,N$, a Weyl orbit of size $N$, the elliptic CM spectral curve, $\S_N$, is
\begin{align}\label{DPcurve}
    \S_N = \bigl\{ \det(y\textbf{1}-L(z))=0 \bigr\},
\end{align}
where $y$ is the sheet coordinate for the spectral curve, $z$ parametrizes an elliptic curve of modulus $\t$ (the position space for elliptic CM systems), and $L$ is the $N\times N$ Lax matrix, 
\begin{align}\label{laxl}
    L(z)=p\cdot H + O(g) .
\end{align}
Here $p\in\C^r$ is the vector of canonical momenta, $H$ is an $r$-component vector of $N\times N$ diagonal matrices with the components of the $w_i$ as diagonal entries.
In particular, $H$ is independent of $z$.%
\footnote{In representation theoretic terms, $H$ is a vector of Cartan generators of the Lie algebra in an $N$-dimensional representation, and the $w_i$ are its weight vectors.}

The elliptic CM modulus, $\t$, is identified with the marginal parameter (gauge coupling constant) of the \Nts\ theory.
The $p$ are identified with special coordinates on the CB.
The coupling constants for the integrable system, $g$, are identified with the mass parameters of the field theory.
For \Nf\ SCFTs we thus set $g=0$, in which case the spectral curve \eqref{DPcurve} factorizes as
\begin{align}\label{spec crv}
    \S_N = \Bigl\{
    0 = \prod_{i=1}^N (y - p\cdot w_i) \Bigr\},
\end{align}
which describes (at a generic point, $p$, of the CB) the Riemann surface given by the disjoint union of $N$ copies of an elliptic curve, $E_\t$, of fixed modulus $\t$.
The Jacobian variety of this curve, ${\rm Jac}(\S_N)$, is simply the direct product
\begin{align}
    {\rm Jac}(\S_N) = E_\t \times \cdots \times E_\t
\end{align} 
with $N$ factors.
Its 1-homology inherits a canonical symplectic pairing (a.k.a., principal polarization) 
\begin{align}\label{can sp}
    J_N=\bpm 0 & \textbf{1}_N \\ -\textbf{1}_N & 0 \epm
\end{align}
from the intersection form on $\S_N$.

The Weyl group acts on the Jacobian variety via its action on the $w_i$, that is, by a set of permutations of the $E_\t$ factors in \eqref{spec crv}. 
The task is then to identify a rank-$r$ (i.e., $r$-complex-dimensional) sub-Abelian variety $X_r \subset {\rm Jac}(\S_N)$ which is invariant under this Weyl group action.
If $z = (z^i)\in\C^N$ is an $N$-component vector of flat coordinates on the $N$ $E_\t$ factors in ${\rm Jac}(\S_N)$, then $X_r$ is specified by an $r$-dimensional subspace $\C^r \subset \C^N$ which is invariant under the action of the Weyl group.
Denote by $P: \C^N \to \C^r$ the projection operator onto this subspace.
The 1-homology of ${\rm Jac}(\S_N)$ is a rank-$2N$ lattice in $\C^N$, and projects to a rank-$2r$ lattice in $\C^r$, which is the EM charge lattice on the CB.
The symplectic pairing \eqref{can sp} projects to a symplectic pairing on the charge lattice, which is the Dirac pairing.
The diagonal modulus $\t \textbf{1}_N$ of ${\rm Jac}(\S_N)$ interpreted as a symmetric bilinear form on $\C^N$ projects to a symmetric form on $\C^r$, which is the matrix of low energy couplings on the CB. 
Finally, the SW 1-form is the canonical 1-form of the elliptic CM system, which is the projection of $\L_N = p\cdot w_i dz^i$, to $\C^r$. 

We can think of the set of $N$ points, $\{w_i\}$, as basis vectors of an $N$-dimensional vector space equivalent to $\C^N$.  
Then the permutation action of the Weyl group on this set of basis vectors defines an $N$-dimensional representation, $O: {\rm Weyl} \to \C^N$, of the Weyl group.
The decomposition of this representation into irreducible components,
\begin{align}
O &\simeq \bigoplus_{j=1} o_j {\bf R_j},
\end{align}
(where ${\bf R_j}$ denote inequivalent irreducible representations of the Weyl group and $o_i$ are their multiplicities) then gives a decomposition of $\C^N$ into Weyl-invariant subspaces.
We want to identify $\C^r$ as an $r$-dimensional invariant subspace.

In general, there can be many such inequivalent subspaces; indeed, even for the $BC_2$ and $G_2$ spectral curves computed in the literature there are 2 and 3 inequivalent $r=2$ invariant subspaces, respectively.
So the condition of Weyl invariance is insufficient to specify a unique CB geometry from the spectral curve.
There is, however, a stronger criterion than mere Weyl invariance that uniquely determines the rank-$r$ invariant subspace, and hence the CB geometry.
To the best of our knowledge, this criterion has not been pointed out previously.
The stronger criterion is that the invariant subspace must be the one corresponding to the irreducible $r$-dimensional reflection (or defining) representation of the Weyl group.
This appears uniquely in the decomposition of $D$ simply because $D$ came from the Weyl group reflection action on a point in $\C^r$.

The physical reason this is the correct prescription is that the reflection representation of the Weyl group gives the orbifold action describing the CB, which in turn gives the EM monodromies acting on the special coordinates, the charge lattice, and as automorphism twists of the SW curve, as explained in section \ref{sec3}.
Thus, in order for the spectral curve to reproduce the correct CB EM monodromies, the Weyl group must act on the invariant subspace via the irreducible reflection representation.

We will now use this criterion to analyze the CB special Kahler structures predicted by the spectral curves derived in the literature for the $G_2$ and $BC_2$ elliptic CM systems.

\paragraph{G$_2$ spectral curve.}

A Lax pair for the $G_2$ elliptic CM system was constructed in \cite{Bordner:1998sw, Bordner:1998xs, Bordner:1998xsa}, and the resulting spectral curve has the form \eqref{DPcurve} corresponding to a (union of) Weyl orbits of $N=6$ points.
Weyl$(G_2)$ is the dihedral group $D_6$ of symmetries of a regular hexagon, and the $N=6$ orbit can be taken as the orbit of the 6 vertices of the hexagon.
The resulting 6-dimensional permutation representation, $O$, is then easily seen to decompose into a sum of four irreducible representations, $O={\bf 1} \oplus {\bf 1_A} \oplus {\bf 2_S} \oplus {\bf 2_R}$, where the last is the reflection representation.
If we label the points $w_i$, $i,\ldots,6$, in the orbit as the vertices going counterclockwise around the hexagon, then the ${\bf 2_R}$ invariant subspace is the one spanned by $\langle w_1{-}w_3{-}w_4{+}w_6, \, w_1{+}w_2{-}w_4{-}w_5 \rangle$.
It then immediately follows that the matrix of low energy couplings on the CB is fixed to $\t \bspm 2&-1\\-1&2\espm$ which is the expected result, but the induced Dirac pairing is
\begin{align}
    J 
    = \bpm &&\ph{-}2&-1 \\ &&-1&\ph{-}2\\ -2&\ph{-}1&&\\ \ph{-}1&-2&&\epm
    \simeq_\Z \bpm &&\ph{-}1& \\ &&&\ph{-}3\\ -1&&&\\ &-3&&\epm,
\end{align}
which is not principal.

\paragraph{BC$_2$ spectral curve.}

The $BC_2$ elliptic CM spectral curve constructed in \cite{DHoker:1997hut, DHoker:1998rfc, DHoker:1998xad, DHoker:1998zuv} is of the form \eqref{DPcurve} corresponding to a Weyl orbit of $N=4$ points.
Weyl$(BC_2)$ is the dihedral group $D_4$ of symmetries of a square, and the $N=4$ orbit can be taken as the orbit of the 4 corners of the square.
The resulting 4-dimensional permutation representation, $O$, is then easily seen to decompose into a sum of three irreducible representations, $O={\bf 1} \oplus {\bf 1_A} \oplus {\bf 2_R}$, where the last is the reflection representation.
If we label the points $w_i$, $i,\ldots,4$, in the orbit as the vertices going counterclockwise around the hexagon, then the ${\bf 2_R}$ invariant subspace is the one spanned by $\langle w_1{-}w_3, \, w_2{-}w_4 \rangle$.
It then immediately follows that the induced Dirac pairing is principal, but the matrix of low energy couplings on the CB is fixed to $\t \bspm 1&0\\0&1\espm$ which is the split form.

\acknowledgments

It is a pleasure to thank A. Bourget, O.Chalykh, J. Grimminger, M. Lotito, Y. Lu, R. Moscrop, S. Thakur, and M. Weaver for helpful conversations.  
PCA is supported in part by DOE grant DE-SC1019775. 
ZY would like to thank University of Chinese Academy of Sciences for support during the master program. MM is supported in part by the STFC through grant number ST/X000753/1.

\nocite{*}
\bibliographystyle{jhep}

\end{document}